\documentclass[sigconf, screen, 10p, anonymous=False]{acmart}
\renewcommand\footnotetextcopyrightpermission[1]{}
\PassOptionsToPackage{dvipsnames}{xcolor}

\PassOptionsToPackage{colorlinks=true,linkcolor=RubineRed,breaklinks=True,citecolor=RubineRed}{xcolor}
\usepackage{enumitem}
\setlist[itemize]{noitemsep, topsep=0pt}

\usepackage[most]{tcolorbox}
\usepackage{adjustbox}
\usepackage[linesnumbered,algo2e,ruled]{algorithm2e}
\usepackage{multirow}
\usepackage{booktabs,subcaption,dcolumn}
\usepackage{tikz}
\usepackage{enumitem}
\usepackage{tabularx}

\usepackage{pifont}
\usepackage{svg}
\usepackage{soul}
\usepackage{amsmath}
\usepackage{dirtytalk}
\usepackage{subcaption}
\usepackage{graphicx}
\usepackage{setspace}
\usepackage[font=small]{caption}
\usepackage{booktabs} 
\usepackage[titletoc,toc,title]{appendix}
\usepackage{svg}
\usepackage{amsfonts}
\usepackage{xurl} 

\usepackage{enumitem}
\graphicspath{{Figures/}}
\svgpath{{Figures/}}

\usepackage{amsmath}
\usepackage{amsfonts} 

\usepackage{colortbl}

\newcolumntype{?}{!{\vrule width 1.5pt}}

\newcommand{\summary}[1]{
    \noindent\fbox{%
        \parbox{0.99\columnwidth}{%
            \textbf{Summary}: {#1}
        }%
    }
}

\newcommand\feature[1]{{\footnotesize{\fontfamily{qhv}\selectfont #1}}}

\newcommand\smamath[1]{{\small $#1$}}

\newcommand\res[2]{\small{$#1$}\tiny{\text{$\pm #2$}}}
\newcommand\ftcal[1]{{\footnotesize $\mathcal{#1}$}}
\newcommand\smacal[1]{{\small $\mathcal{#1}$}}
\newcommand\sccal[1]{{\scriptsize $\mathcal{#1}$}}
\newcommand{\overbar}[1]{\mkern 1.5mu\overline{\mkern-1.5mu#1\mkern-1.5mu}\mkern 1.5mu}

\newtcolorbox{cooltextbox}[1][]{%
    colback=black!5,
    colframe=black!5,
    notitle,
    sharp corners,
    borderline west={0pt}{0pt}{red!80!black},
    enhanced,
    breakable,
    left=0pt,
    right=0pt,
    top=0pt,
    bottom=0pt
    }

\newcommand\revision[1]{%
  \bgroup
  \hskip0pt\color{blue!80!black}%
  #1%
  \egroup
}

\newcommand\api[1]{{\fontfamily{pcr}\selectfont {\footnotesize #1}}}

\theoremstyle{definition}

\newcommand\bestres[2]{\small{$\mathbf{#1}$}\tiny{\text{$\mathbf{\pm #2}$}}}

\newcommand\ours[1]{\small{${#1}$}}

\AtBeginDocument{%
  \providecommand\BibTeX{{%
    \normalfont B\kern-0.5em{\scshape i\kern-0.25em b}\kern-0.8em\TeX}}}

\AtBeginDocument{%
  \providecommand\BibTeX{{%
    \normalfont B\kern-0.5em{\scshape i\kern-0.25em b}\kern-0.8em\TeX}}}

\copyrightyear{2023}
\setcopyright{rightsretained}
\acmConference[CODASPY '23]{Thirteenth ACM Conference on Data and Application Security and Privacy}{April 24--26, 2023}{Charlotte, NC, USA}

\begin{document}

\title{Attribute Inference Attacks in Online Multiplayer Video Games: a Case Study on \textsc{Dota2}}


\author{Pier Paolo Tricomi}
\email{tricomi@math.unipd.it}
\orcid{0000-0003-1600-835X}
\affiliation{%
    \department{Department of Mathematics}
  \institution{† University of Padua, Italy}
  \country{}
}

\author{Lisa Facciolo}
\email{lisa.facciolo@studenti.unipd.it}
\orcid{0000-0002-7041-4693}
\affiliation{%
  \department{Department of Mathematics}
  \institution{† University of Padua, Italy}
  \country{}
}

\author{Giovanni Apruzzese}
\orcid{0000-0002-6890-9611}
\email{giovanni.apruzzese@uni.li}
\affiliation{%
  \department{Hilti Chair of Data and Application Security}
  \institution{University of Liechtenstein}
  \country{}
}

\author{Mauro Conti}\authornotemark[2]
\email{conti@unipd.it}
\orcid{0000-0002-3612-1934}
\affiliation{%
  \department{Faculty of EEMCS}
  \institution{Delft University of Technology, NL}
  \country{}
}

\begin{abstract}

Did you know that over 70 million of \textsc{Dota2} players have their in-game data freely accessible? What if such data is used in malicious ways? This paper is the first to investigate such a problem.

Motivated by the widespread popularity of video games, we propose the first threat model for Attribute Inference Attacks (AIA) in the \textsc{Dota2} context. We explain \textit{how} (and \textit{why}) attackers can exploit the abundant public data in the \textsc{Dota2} ecosystem to infer private information about its players. Due to lack of concrete evidence on the efficacy of our AIA, we empirically prove and assess their impact in reality. By conducting an extensive survey on $\sim$500 \textsc{Dota2} players spanning over 26k matches, we verify whether a correlation exists between a player's \textsc{Dota2} activity and their real-life. Then, after finding such a link ($p\!<\!0.01$ and $\rho>0.3$), we ethically perform diverse AIA. We leverage the capabilities of machine learning to infer real-life attributes of the respondents of our survey by using their publicly available in-game data. Our results show that, by applying domain expertise, some AIA can reach up to 98\% precision and over 90\% accuracy.
This paper hence raises the alarm on a subtle, but concrete threat that can potentially affect the entire competitive gaming landscape. We alerted the developers of \textsc{Dota2}.

\end{abstract}



\settopmatter{printfolios=true}

\maketitle

\section{Introduction}

More than 3 billion people played Video Games (VG) in 2021, whose industry is constantly expanding, attracting new players every day~\cite{gamingstat}. 
A recent study highlighted that over 71\% of participants increased their playtime, and that VG improved their well-being~\cite{gamerswellbeing}. Within the broad VG landscape, one category stands out: \textit{online multiplayer VG}. These VG allow players to interact with each other in a `controlled' environment (i.e., the game) that is separated from their private life~\cite{tally2021protect}. Specifically, players can interact in two distinct settings: cooperative or competitive. This paper focuses on the latter, motivated by the rise of the Electronic Sports (E-Sports) panorama, which generated over \$1B of revenues in 2021~\cite{gamingstat}.

In E-Sports, players compete in VG matches~\cite{hamari2017esports}. Notable examples of E-Sports VG are Fortnite, ApexLegends, CS:GO, and \textsc{Dota2}. All such VG are addictive (on average, \textsc{Dota2} players have over 1600 hours of playtime), and have an heterogeneous playerbase~\cite{professionals}. Some individuals ``play for fun'', e.g., to spend their free-time with friends, or to entertain their audience on streaming platforms~\cite{kaytoue2012watch}. Others, however, ``play to win'', and their primary aim is improving so that they can participate in (and, perhaps, win) one of the many competitions held regularly. Such competitions have rich prize-pools (up to \$40M~\cite{TI}) which attract thousands of contestants. Indeed, winning matches is difficult due to the highly competitive environment (which is ultimately a zero-sum game~\cite{guo2021adversarial}), and `mastering' an E-Sport VG requires constant dedication~\cite{griffiths2017psychosocial}.

\begin{figure}[!htbp]
    \centering
    \frame{\includegraphics[width=0.95\columnwidth]{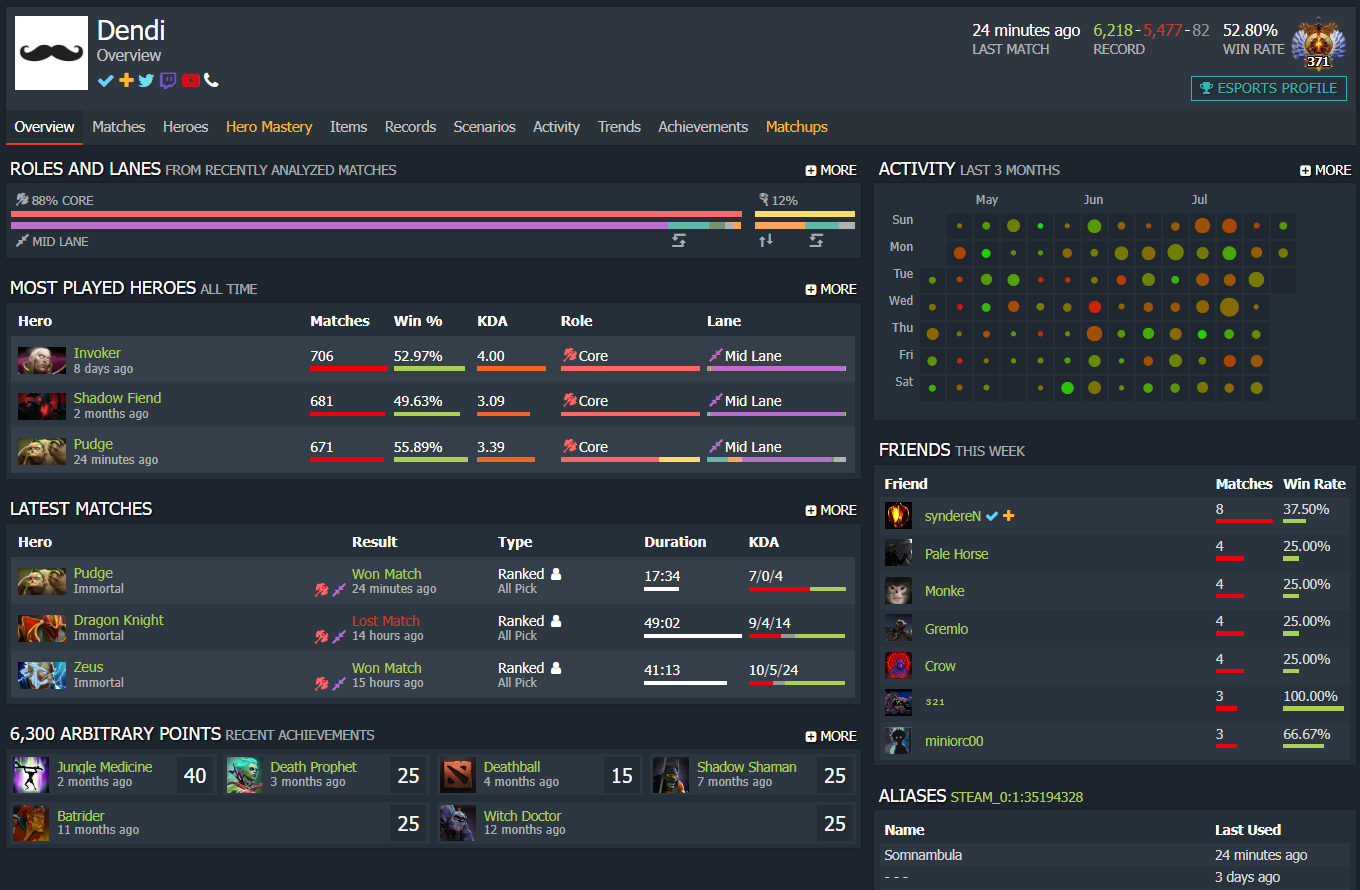}}
    \caption{A TW, showing the statistics of the professional \textsc{Dota2} player ``Dendi''~\cite{dendi}. All such information is constantly updated and publicly accessible: {\small{\url{https://dotabuff.com/players/70388657}}}.}
    \label{fig:screenshot}
    \vspace{-2.5em}
\end{figure}

Several resources, typically referred to as Tracking Websites (TW), were born to track players' activities on a specific VG. Indeed, an intuitive way to improve is learning from past mistakes, and TW greatly facilitate such process by providing their users (i.e., the players) with detailed statistics of their in-game performance. We provide a screenshot of a TW focused on \textsc{Dota2} in Fig.~\ref{fig:screenshot}, showing an overview of the in-game activities of the player ``Dendi''. 
Such statistics include, e.g., data of past matches, the days in which the player is more active, their friends; additional information is available by navigating the webpage. 
Despite the undeniable advantages provided by TW (over 70M of \textsc{Dota2} players use TW~\cite{stratz_pl}), we observe that \textit{all data retrieved and elaborated by TW is publicly available}: anyone can observe, collect, and use such data.

\vspace{-2.5em}

We thus ask ourselves: ``what if players' in-game data are used \textit{against} them to violate their privacy?'' If this were true, then the E-sport setting would be prone to Attribute Inference Attacks (AIA). Such attacks, enabled by the capabilities of Machine Learning (ML), aim to infer private information about a given target (i.e., a player) by using their publicly available data~\cite{gong2016you}.
Although TW report only in-game statistics, we cannot exclude the existence of a link between such data and personal attributes (e.g., gender, age, personality) or even sensitive ones (e.g., health~\cite{sensitiveEU, sensitiveUSA})---the latter being outside our scope. Prior research (e.g.,~\cite{tekofsky2013psyops, martinovic2014you}) revealed that a correlation exists between the in- and off-game traits of a given player. Surprisingly, however, no study has been carried out within the specific context of \textsc{Dota2}. Such a lack is concerning: the in-game data provided by \textsc{Dota2} is semantically different from that of other VG. Hence, to this day, it is still \textit{uncertain whether AIA are a threat} to \textsc{Dota2} players. Consequently, it is also unknown (i)~\textit{how} AIA can be carried out and (ii)~what is the \textit{impact} of an AIA in the \textsc{Dota2} context.
Inspired by Biggio and Roli~\cite{biggio2018wild}, we proactively assess the likelihood and the effects of this subtle privacy issue.


\textsc{\textbf{Contribution.}}
This paper investigates the threat of AIA against \textsc{Dota2} players. 
We begin (§\ref{sec:background}) by contextualizing the E-Sports ecosystem (with a focus on \textsc{Dota2}) and summarizing the fundamental concepts of AIA (building on related work). Then, we provide four major contributions---which go \textit{beyond the research domain}.
\begin{itemize}
    \item \textbf{A threat model of AIA against \textsc{Dota2} players}~(§\ref{sec:threat}). We describe \textit{how} to (legitimately) launch an AIA to infer private information on players while knowing only their \textsc{Dota2} handle. We also explain \textit{why} attackers would do so.
    
    \item \textbf{We prove the existence of correlations between \textsc{Dota2} players' in-game data and their personal attributes}~(§\ref{sec:preliminary}). By conducting an (informed) survey, we collect in-game and personal data of 484 \textsc{Dota2} players, corresponding to over 26k matches. We then perform a correlation analysis, showing the existence of statistically significant ($p<0.01$) and strong (Spearman's $\rho>0.3$) relationships between in-game (public) and off-game (private) attributes. 
    
    \item \textbf{We proactively evaluate the impact of AIA in \textsc{Dota2}}~(§\ref{sec:evaluation}). We use the data gathered from our survey to (ethically) enact an AIA, and measure its success rate. We develop multiple ML models, by assuming attackers with varying domain expertise on \textsc{Dota2}. We show that even simple AIA can be successful (almost 70\% F1-score on \feature{gender}), and that more sophisticated AIA can further increase such impact (over 75\% accuracy on predicting the \feature{occupation}).
    
    \item \textbf{We assess AIA that can be staged in practice}~(§\ref{sec:practical}). 
    We assume the viewpoint of an attacker with \textit{specific} goals, and elucidate the real-threat of AIA in \textsc{Dota2} by demonstrating a realistic application of our findings, showing AIA with near-perfect success rate (almost 100\% precision).
\end{itemize}
Finally, we discuss our results, describe some countermeasures, and explain how our AIA can be extended to other E-Sports VG (§\ref{sec:discussion}). We then conclude our paper and provide ethical considerations (§\ref{sec:conclusion}). 

\textbf{Transparency.} We release a repository containing exhaustive details on our study, as well as the source code we developed for our analyses---available at: \url{https://github.com/hihey54/Dota2AIA}. Finally, we remark that Pier Paolo Tricomi is a top-1\% \textsc{Dota2} player.

\begin{cooltextbox}
\textsc{\textbf{Disclaimer.}}
Our paper tackles a delicate privacy issue that potentially touches millions of video-gamers. All our evaluations are conducted ethically~\cite{spring2017practicing}, but attackers are not bound to such ethics. At the time of writing, the problem is still open. 
\end{cooltextbox}
\section{Background and Related Work}
\label{sec:background}
Our paper tackles two emerging domains: competitive video games, and attribute inference attacks---which we now summarize. 

\subsection{The Competitive Video-Game Ecosystem}
\label{ssec:ecosystem}
\begin{figure}[!htbp]
    \centering
    \includegraphics[width=0.6\linewidth]{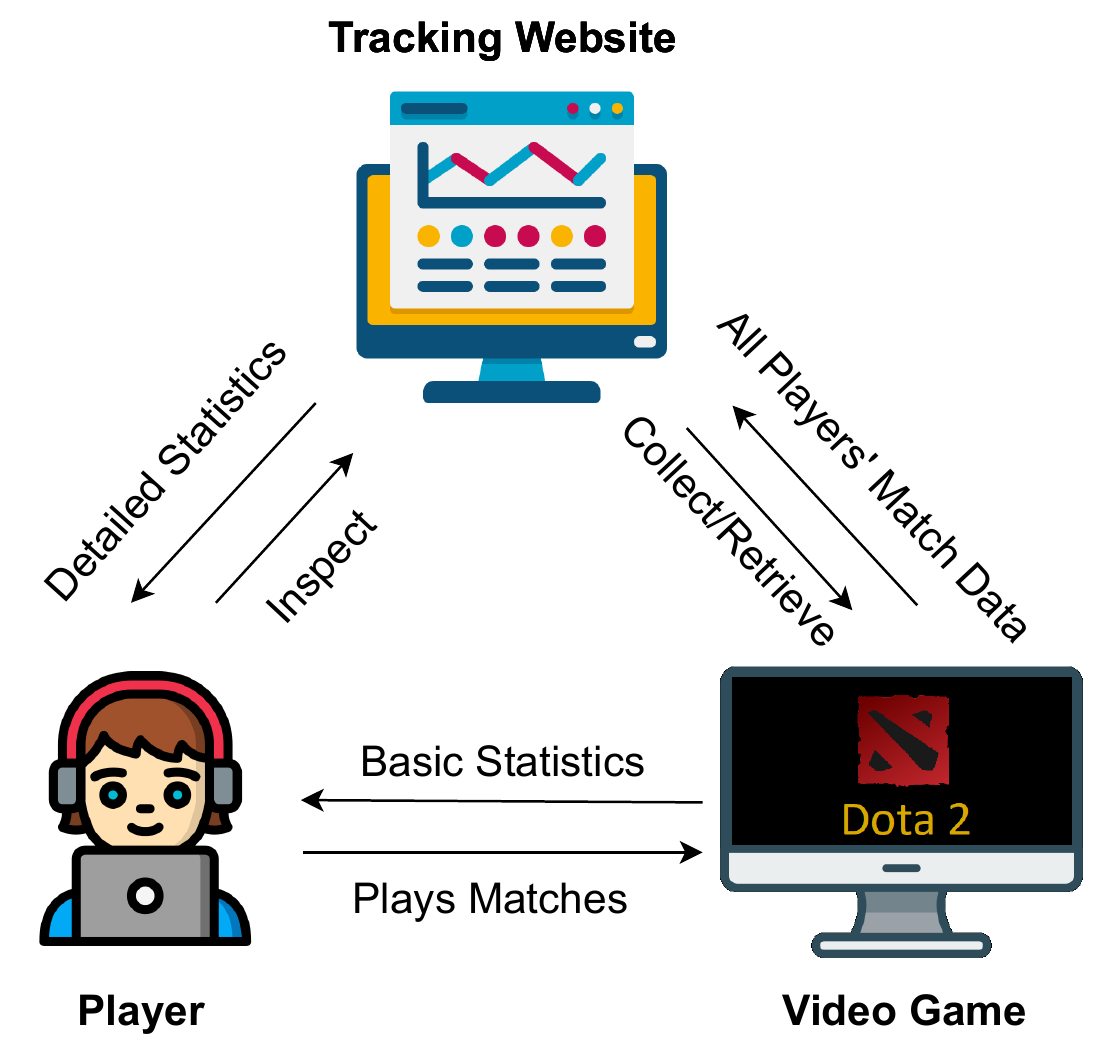}
    \vspace{-3mm}
    \caption{The E-Sport ecosystem. \textit{Players} engage in \textit{matches} of a \textit{video-game}, which publicly releases data on such matches. These data are collected by \textit{tracking websites}, whose elaborations are made public.}
    \label{fig:ecosystem}
    \vspace{-3mm}
\end{figure}

Competitive \textit{video-games} (VG), and E-Sports in particular, are receiving a lot of attention~\cite{gamingstat}, leading to a constant increase of players all aiming to ``reach the top''~\cite{karkallis2021detecting}. To improve their performance, players can analyze their in-game statistics~\cite{improve}. Such statistics are typically provided by the VG itself, but are limited to a single \textit{match}. Even if most VG allow players to inspect their history, analyses can only be performed on a match-by-match basis. Such limitation was overcome by \textit{Tracking Websites} (TW), which collect and aggregate information pertaining to all matches of a given player(s), providing a comprehensive overview of their activity (cf. Fig.~\ref{fig:screenshot}). An illustration of such ecosystem is in Fig.~\ref{fig:ecosystem}, which we now describe from the viewpoint of our VG of choice---\textsc{Dota2}. 

\vspace{-0.5em}

\begin{itemize}
    \item \textbf{Video-Game.} 
    \textsc{Dota2} is a Multiplayer Online Battle Arena (MOBA) VG. Released in 2013 and available for free, it is one of the most popular VG, counting up to 6M daily players and over 15M monthly players~\cite{activeplayer}. In a match, two teams of five players fight in real time with a common objective: destroy the enemy team's base before they do it to yours.
    
    \item \textbf{Players.} 
    Each player in a team has a crucial role in ensuring their team's victory, and such roles are difficult to master.
    Indeed, \textsc{Dota2} is extremely competitive: in 2021, its biggest tournament had the largest prize pool in the entire history of VG, amounting to \$40M~\cite{TI}. Such prizes are enticing for players, who continuously strive to get better: every \textsc{Dota2} player has more than 1600 hours~\cite{dota_playtime} of playtime (on average). It is not surprising, hence, that \textsc{Dota2} players will resort to any (legitimate) tool to maximize their efficiency. 
    
    \item \textbf{Tracking Websites.} A massive amount of \textsc{\textsc{Dota2}} players leverage the services provided by TW~\cite{conti2020pvp}. Reportedly, some  TW tracked the activities  of more than 79M players, aggregating the results of $\sim$3B matches~\cite{stratz_pl}. In our context, TW constantly interact with specific \textsc{Dota2} APIs to retrieve all historical data pertaining to a player's matches. Before using a TW, a player must explicitly allow \textsc{Dota2} to share their match details with external sources; however, considering the benefits provided by TW, only few players do not give their consent. Every player (and corresponding \textsc{Dota2} activity) tracked by a TW is publicly visible on the platform.
\end{itemize}

\noindent
Such context begs the question: ``\textbf{why are TW publicly releasing players' data?}'' The answer is: ``because players themselves want such data to be public.'' Indeed, such availability allows players to:
\begin{itemize}
    \item browse other players' statistics, so as to learn how the game is played by top-players;
    \item increase their visibility to professional organizations, which can hire them if they show good performance;
    \item share their activity with friends, teammates, or even unknown players that paired up with them;
    \item climb TW-specific rankings (e.g., players who get most wins with a given character).
\end{itemize}
Simply put, players benefit from their in-game data being publicly released by TW---thereby exposing players to the threat of AIA. 

\subsection{Attribute Inference Attacks}
\label{ssec:attribute}
We summarize the fundamentals of Attribute Inference Attacks (AIA), and then highlight the research gap motivating our paper.

\subsubsection{AIA in a nutshell}
\label{sssec:aia}
The underlying goal of AIA is inferring \textit{private} information on a given target by exploiting \textit{publicly available} data on such target. 
For example, an attacker can use the (public) ratings posted on a video streaming platform by a given user to infer their (private) gender~\cite{weinsberg2012blurme}. Such inference can be done leveraging the predictive capabilities of Machine Learning (ML): By training a ML model on a representative dataset, and then providing such ML model with some user's public data, the ML model will output the personal attributes of such user. 
We remark that AIA are semantically different than membership inference attacks (e.g.,~\cite{jarin2021pricure, zhong2022understanding}), whose goal is inferring information on the ML model's training set.

AIA are becoming problematic due to the lack of education of most internet users, who publicly share their data while overlooking (or ignoring) the corresponding risks (e.g.,~\cite{cheng2013preserving, ilia2015face, morris2021you}). For instance, most data published on social networks can be easily retrieved via OSINT~\cite{apruzzese2022role} and then used to setup an AIA. Indeed, most prior research considers the ecosystem of social networks, due to the ease of retrieving information linking public data with private attributes: Goelbeck et al.~\cite{golbeck2011predicting} infer personality traits of social media users.
Jurgens et al~\cite{jurgens2015geolocation} consider Twitter, and predict the location of the users based on their tweets.
More recently, Gong et al.~\cite{gong2018attribute} focus on Google+ users, whereas Zhang et al.~\cite{zhang2020practical} consider, e.g., YouTube, and predict users' gender (above 70\% F1-score) based on their historical activity. Similarly,~\cite{pijani2020you} focus on Facebook, showing that the gender can be predicted ($\sim$80\% accuracy) by analyzing the usage of emojis. (The authors of~\cite{kosinski2013private} also consider Facebook, and infer sensitive data which is outside our scope). Other examples are~\cite{chen2014effectiveness, weinsberg2012blurme, yo2017inference, eidizadehakhcheloo2021divide}. All such works show that AIA can be enacted in the real world, representing a subtle privacy risk.


\subsubsection{Motivation: AIA and Video Games}
\label{sssec:related}
Surprisingly, no efforts consider AIA exploiting (public) VG data to infer players' (private) attributes---to the best of our knowledge. As shown in §\ref{ssec:ecosystem}, the competitive VG ecosystem (and especially the one of \textsc{Dota2}) is particularly prone to the risk of AIA. A trace of such exposure is provided by the few works analyzing the correlation between the players' in-game behaviour and their personal life---albeit for VG of different genres. For instance, Oggins et al.~\cite{oggins2012notions} highlighted that 
MMORPG players have a similar in- and off-game behaviour. 
Martinovic et al.~\cite{martinovic2014you} reasoned on how such similarity can be used by producers of MMORPG. For instance, some players' physical traits can be inferred from their in-game avatar---which tends to be alike~\cite{nowak2005influence}. In this context, Likarish et al.~\cite{likarish2011demographic} analyzed the in-game avatars to predict the age of the corresponding player; whereas Symborski et al.~\cite{symborski2014use} predicted the gender. Besides physical characteristics, some researches also studied personality indicators. Spronck et al.~\cite{spronck2012player} found correlations between personality traits of 36 players and their playing-style. The only paper we are aware of that considers a competitive VG is~\cite{tekofsky2013psyops}, showing correlations between Battefield3 players' in-game data and some of their personality traits. 

Most related studies on VG (i) did not consider MOBA---which are our focus; and (ii) adopted the perspective of the producers of the VG---i.e., they assumed the availability of in-game data that was not publicly available~\cite{sifa2018profiling, drachen2014comparison}. The latter is crucial: a \textit{real} attacker is unlikely~\cite{apruzzese2021modeling} to have access to a company's databases---especially in domains with a high market share, such as  (competitive) VG. Granted: such studies showed that correlations exist between players' in- and off-game characteristics, \textit{but in different VG}. No paper, however, investigated: (i)~whether a correlation exists also in \textsc{Dota2}; and, if it exists, (ii)~`how' and `to what extent' it can be exploited in the \textsc{Dota2} ecosystem by real attackers---who are not omnipotent. 
The only work that considers a similar setting as ours is~\cite{conti2020pvp}, but it focused on recognizing the play-style of \textsc{Dota2} players across different accounts---which is an objective orthogonal to ours.
To the best of our knowledge, we are the first to investigate AIA in VG.

\section{\textsc{Dota2} Attribute Inference Attacks}
\label{sec:threat}
Our primary contribution is the first threat model for feasible AIA against \textsc{Dota2} players. We describe `how' AIA can be staged in the \textsc{Dota2} ecosystem (§\ref{ssec:definition}); and `why' attackers would do so (§\ref{ssec:feasibility}).

\subsection{Proposed Threat Model}
\label{ssec:definition}
Our AIA is mostly tailored for players who actively engage in \textit{competitive} \textsc{Dota2} matches. (Some \textsc{Dota2} players do not ``play to win'', and hence are less likely to use TW.) For simplicity, we assume that a player only owns a single `handle' (e.g., ``Dendi'' in Fig.~\ref{fig:screenshot} is the handle of the player ``Danil Ishutin''), which is used to retrieve data from any public source (e.g., tracking websites). 

\textbf{Formal Definition.}
We describe the viewpoint of our considered attacker according to the following four criteria~\cite{biggio2018wild}:
\begin{itemize}
    \item \textit{Goal:} The attacker wants to infer the personal attributes of a set of players whose real identity is completely private.
    
    \item \textit{Knowledge:} The attacker knows the handles of a set of players, and is well-aware of the \textsc{Dota2} ecosystem. 
    
    \item \textit{Capability:} The attacker can only access and retrieve data that is either publicly available, or that users are willing to share (e.g., social networks, public surveys). 
    
    \item \textit{Strategy:} The attacker first (legitimately) gathers information associating players' in-game data with their respective personal attributes. Then, the attacker trains a ML model to perform AIA against players whose personal information is completely private, i.e., by only using their (known) handle.
\end{itemize}
We implicitly assume that the targeted players enabled in-game data sharing with external sources (e.g., TW). We stress that the attacker shall \textit{not} perform any data breach to obtain the desired private information---an attacker will never launch an AIA otherwise.

\begin{figure}[!htbp]
    \centering
    \includegraphics[width=0.93\columnwidth]{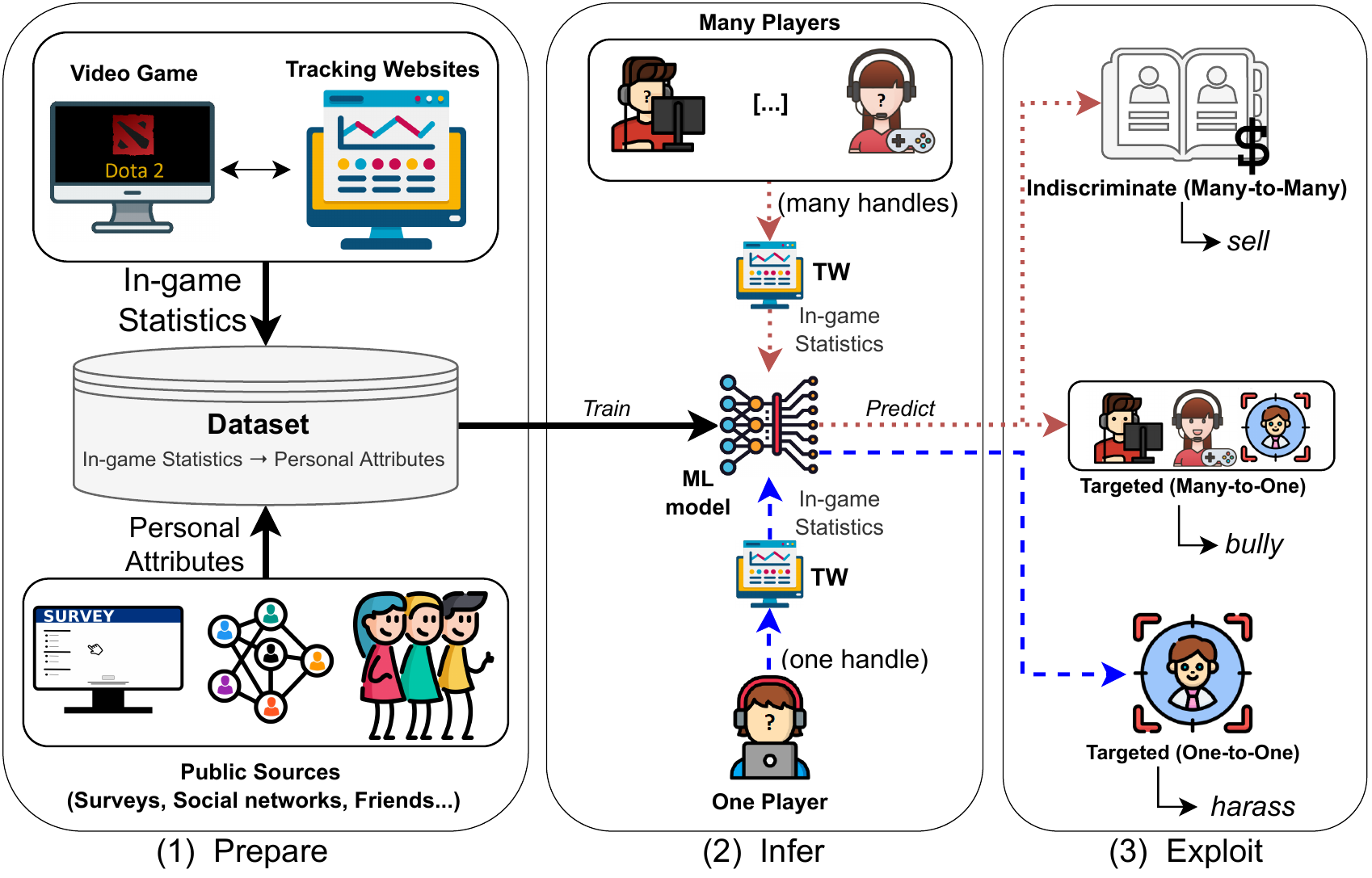}
    \caption{Overview of our proposed AIA against \textsc{Dota2} players. Public information is used to infer personal (private) attributes. Besides privacy violations, attackers can harass or bully their victims, or profit from all the inferred attributes.}
    \label{fig:over_att}
    
\end{figure}

\textbf{Practical scenario.} 
We present in Fig.~\ref{fig:over_att} an illustration of our threat model, which is divided in three stages: \textit{prepare}, \textit{infer}, \textit{exploit}.

\begin{enumerate}
    \item \textit{Prepare.} First (left of Fig.~\ref{fig:over_att}), the attacker must collect a representative dataset associating \textsc{Dota2} players' in-game data (e.g., daily matches played, win/loss ratio) with the corresponding ground truth (e.g., the players' gender).
    
    \item \textit{Infer.} Then (middle of Fig.~\ref{fig:over_att}), the attacker uses the harvested dataset to train a ML model, which is the tool to carry out the AIA. The inference is done by providing public in-game information on a target player (obtainable, e.g., by querying a TW with the handle of a player) as input to the ML model.
    
    \item \textit{Exploit.} 
    Finally (right of Fig.~\ref{fig:over_att}), the attacker benefits by either stalking a victim (targeted AIA), or by profiting from the inferred attributes (an indiscriminate AIA). 
\end{enumerate}

\vspace{-1em}

\subsection{Feasibility of AIA in \textsc{Dota2}} 
\label{ssec:feasibility}
Any attack is theoretically possible, and several papers (e.g.,~\cite{arp2022and}) advocate to always consider worst-case scenarios. Nonetheless, we argue that our proposed AIA are not only possible ``in theory'', but also likely to occur ``in practice'' due to their high feasibility~\cite{apruzzese2021modeling}. Indeed, real attackers have a cost-benefit mindset~\cite{wilson2014some}. In our case, AIA will be launched only if an attacker finds them easy to setup (in terms of cost and risk), and if they lead to tangible benefits.

In particular, we focus the attention on three aspects---each pertaining to a given stage of our exemplary use-case, namely: acquiring the dataset to train the ML model (i.e., the \textit{capabilities} of the attacker); improving the performance of such ML model (i.e., the \textit{knowledge} of the attacker); and how a successful AIA can be exploited (i.e., the \textit{goal} of the attacker).\footnote{We observe that our threat model is significantly different from the one in~\cite{mehnaz2022your}.}

\begin{itemize}
    \item \textbf{Data Harvesting.} Obtaining \textit{public} in-game data of multiple players (i.e., the ``features'') is straightforward in \textsc{Dota2}: it is simply necessary to go to a TW\footnote{We observe that abundant information is also available directly from \textsc{Dota2}, hence TW are not strictly required (we will discuss this in §\ref{sec:discussion}).} and retrieve all information related to a set of players. In contrast, obtaining the corresponding personal attributes (i.e., the ``labels'') may appear harder, as such information is typically kept \textit{private}. Unfortunately, this is not the case in the \textsc{Dota2} ecosystem.\footnote{Zhang et al.~\cite{zhang2020practical} also state that ground-truth harvesting is easy in today's landscape.} For instance, the real identity of many players (e.g., professionals or streamers) is well-known. Moreover, it is possible to search for a given handle on popular search-engines and inspect the results. For example, a given player may use the same handle also on social media; some people even announce their handle on public forums to facilitate establishment of partnerships. Alternatively it is also possible to conduct surveys in which interviewees must input their handle, as well as some inconspicuous private information (e.g., gender, age). For instance, two large surveys were carried out in 2016 and 2021, \textit{receiving 30k and 8k responses} respectively, by simply posting announcements on popular boards~\cite{surveyreddit}. 
    
    \item \textbf{Refining the ML model performance.} Even if an attacker can acquire a suitable training dataset, it is unlikely that such dataset can yield a proficient ML model from the start---hence, \textit{naive} attackers will hardly be successful in their AIA. \textit{Expert} attackers, however, can use their superior knowledge on the \textsc{Dota2} scene to improve the success rate of their AIA. In our evaluation (§\ref{sec:evaluation}) we will show some pre- and post-processing techniques that boost the predictive performance of the ML model. Given that attackers interested in our AIA are well-aware of how \textsc{Dota2} works, this characteristic further aggravates the threat of AIA.
    
    \item \textbf{Exploiting AIA}. We identify three ways in which an attacker can benefit from AIA in \textsc{Dota2}. (We will consider all of these ways in our evaluation.) First, they can launch an \textit{indiscriminate} `many-to-many' AIA, i.e., by using many handles (belonging to many players) to infer the respective personal attributes; such attributes can then be sold\footnote{This is a popular strategy adopted by some real companies~\cite{mohajeri2019watching}.} to any potential buyer---e.g., dark web, or even to ad-companies which want to send customized ads~\cite{schneier2015data}). Second, they can launch a \textit{targeted} `one-to-one' AIA by inferring the attributes of just one player---e.g., after losing a match, an attacker can launch an AIA against a player of the opposing team and harass them~\cite{conti2020pvp}). Third, they can launch a \textit{targeted} `many-to-one' AIA by inferring the attributes of a set of (many) players, and then finding a (single) player within such set that meets some criteria---e.g., finding an underage player and then bully them~\cite{fryling2015cyberbullying, groomingLol, groomingFort}. 
\end{itemize}
Finally, we observe that the results of the two surveys~\cite{surveyreddit} showed similar trends despite the 5 year timespan. Such stability may suggest that even \textbf{data collected many years prior can still be used to enact successful AIA}. Considering the high likelihood of such a threat, we embrace Biggio and Roli's~\cite{biggio2018wild} recommendation: we must proactively assess the impact of AIA in \textsc{Dota2}.

\begin{cooltextbox}
\textsc{\textbf{Takeaway:}} Attackers can -- cheaply and legitimately -- use many methods to setup an AIA, which can be exploited in various ways to violate \textsc{Dota2} players' privacy.
\end{cooltextbox}

\section{Preliminary Assessment} 
\label{sec:preliminary}
A prerequisite for a successful AIA is the existence of relationships between the players' in-game data, and their corresponding personal attributes~\cite{zhang2021graph}. We recall (§\ref{sssec:related}) that past research found some correlations---but \textit{in different VG} (e.g., Battlefield3~\cite{tekofsky2013psyops}). 

Hence, as our second contribution, we now investigate whether there is some evidence hinting that AIA ``can be successful in \textsc{Dota2}''. To this purpose, we perform an extensive survey on real \textsc{Dota2} players (§\ref{ssec:survey} and §\ref{ssec:ingame}), and analyze the correlation coefficient between their in-game statistics and personal attributes (§\ref{ssec:correlation}).

\subsection{Collection of personal attributes (survey)}
\label{ssec:survey}
We conduct a survey to collect the handles of  \textsc{Dota2} players, together with their personal attributes. 

\textbf{Method.} 
The handle consists in the Steam ID of each player. For the personal attributes, we consider: \feature{gender}, \feature{age}, \feature{occupation}, \feature{purchase\_habits}, as well as the ``Big Five'' personality traits~\cite{wiggins1996five}).
Such attributes are those typically envisioned by past research (e.g.,~\cite{golbeck2011predicting, zhang2020practical, pijani2020you, yo2017inference}); the only exception is \feature{purchase\_habits}, which is an `original' attribute that we propose due to the given \textsc{Dota2} context, in which players typically purchase ``cosmetics'' to embellish their characters. Nevertheless, all such personal attributes represent information that is \textit{not available} from any resource linked with \textsc{Dota2}: hence inferring such information without the explicit consent of the corresponding player represents a privacy violation.\footnote{Even \feature{{\scriptsize purchase\_habits}} is not public: a player may have many ``cosmetics'', which can have been \textit{gifted}; moreover, a single purchase may include \textit{more} than a single ``cosmetic'', which can also be obtained via ``bundles''.}
Our survey entailed 10 questions used to determine the personality traits~\cite{personality}; 4 questions which explicitly referred to the remaining four attributes considered in this paper; as well as one question for the country. We also included 10 questions, which served both as `attention checks', but also for verifying the authenticity of the answers (e.g., we asked ``what is your favorite \textsc{Dota2} hero?'' and we verified on a TW whether the answer was genuine).\footnote{Our repository includes the full questionnaire. Some questions found therein asked for other (non-sensitive) information that do not pertain to this paper.}
Overall, the survey began in Oct. 2019 and ended in Dec. 2019. In this timeframe, we hosted our survey on a website, whose link was distributed on many online social media platforms such as Facebook, Reddit, Discord, and Telegram. Upon landing on the survey's website, participants had to login with their Steam account (via OpenID), thereby ensuring that all personal attributes were correctly linked to the actual player. 

\begin{table}[ht!]
    \centering
    \vspace{-0.5em}
    \caption{Personal attributes considered in our study. Our population is of 484 \textsc{Dota2} players. The distribution resembles the one in~\cite{surveyreddit}.}
    \label{tab:attributes}
    \resizebox{1\columnwidth}{!}{%
        \begin{tabular}{lll}
            \toprule
            \textit{\textbf{Private Attribute}} & \textit{\textbf{Description}} & \textit{\textbf{Classes Distribution}} \\\midrule
            \feature{gender} & Gender at birth & \textit{Female}: (4.96\%), \textit{Male}: (95.04\%) \\
            \feature{age} & Current age & \textit{13--18}: (13.43\%), \textit{19--24}: (53.72\%), \textit{25--38}: (32.85\%) \\
            \feature{occupation} & Whether a player is employed or not & \textit{No}: (57.44\%), \textit{Yes}: (42.56\%) \\
            \feature{purchase\_habits} & Frequency of in-game purchases & \textit{Never}: (10.54\%), \textit{Rarely}: (61.16\%), \textit{Regularly}: (28.30\%) \\
            \feature{openness} & Inventive/curious (high) vs. consistent/cautious (low) & \textit{Low:} (19.22\%), \textit{Medium}: (24.38\%), \textit{High:} (56.40\%) \\
            \feature{conscientiousness} & Efficient/organized (high) vs extravagant/careless (low) & \textit{Low}: (39.46\%), \textit{Medium}: (23.97\%), \textit{High}: (36.57\%) \\
            \feature{extraversion} & Outgoing/energetic (high) vs. solitary/reserved (low) & \textit{Low}: (47.31\%), \textit{Medium}: (21.07\%), \textit{High}: (31.62\%) \\
            \feature{agreeableness} & Friendly/compassionate (high) vs. critical/rational (low) & \textit{Low}: (20.87\%), \textit{Medium}: (19.42\%), \textit{High}: (59.71\%) \\
            \feature{neuroticism} & Sensitive/nervous (high) vs. resilient/confident (low) & \textit{Low}: (53.51\%), \textit{Medium}: (19.21\%), \textit{High}: (27.27\%)\\\bottomrule
        \end{tabular}
    }
    \vspace{-1em}
\end{table}

\textbf{Analysis.}
We received 625 answers from 62 different countries. We filtered out: 18 invalid answers; 43 participants who were not visible on any TW; and 78 inactive players (i.e., less than 5 games in the last month). Thus, our sample size consists in 484 players. Despite being far smaller than the overall amount of \textsc{Dota2} players, such number still allows to draw statistically significant result. Indeed, we are above the minimum sample size of 384 required by setting a confidence level of 95\%, a margin of error of 5\%, population proportion of 50\%, and a population size of 7 million~\cite{kotrlik2001organizational}. 
We report in Table~\ref{tab:attributes} the considered personal attributes, as well as their class-distribution in our population. We grouped \feature{age} in three bins (similarly to~\cite{chen2014effectiveness}): very young/underage, young adults, and over 25 (our `oldest' respondent was 38); the frequencies for \feature{purchase$\_$habits} are never, less than once a month (rarely), and monthly or more often (regularly); for \feature{occupation}, we consider a student as unemployed. Since our survey quantified each personality trait as an integer [0--100], we group such values into three categories (similarly to~\cite{bunian2017modeling}) differentiating low, middle, or high scores.

\textbf{Validation.}
By observing Table~\ref{tab:attributes}, we can see that some classes may present a high imbalance, such as \feature{gender} or \feature{age}. However, our class-distribution is strikingly similar to those of the surveys carried out in previous years~\cite{surveyreddit}: specifically, we focus on the largest survey from 2016, whose sample size was of 29,351. Let us make some exemplary comparisons, so as to \textit{validate} all our subsequent analyses: if our population significantly differs from the `real' one, then we cannot claim that the threat is `real'.
According to~\cite{surveyreddit}, \textit{male} players are $96\%$, which match our results of $95\%$. The same can be said for \feature{age}: according to~\cite{surveyreddit}, minors represent $15\%$ of the population (ours is $13.4\%$), whereas young adults are $66\%$ (ours is $54\%$), with over 25 being $20\%$ (ours is $33\%$). (Small differences are due to slightly different thresholds for the bins). For \feature{occupation}, the unemployed are $67\%$ in~\cite{surveyreddit} (ours is $57\%$).

\summary{from our survey, we derive that: our population (i)~is representative of the \textsc{Dota2} community, and (ii)~is large enough to derive statistically significant conclusions. Moreover, our survey also shows that (iii) the \textsc{Dota2} community is willing to participate in online surveys---representing one of the means an attacker can use to harvest players' private information for a (real) AIA.}

We will use \smacal{A} to indicate the dataset containing the (personal) \textit{attributes} of our 484 players---collected via our ethical survey.\footnote{We never attempt at inferring additional (private) information of our respondents.}

\subsection{Collection of in-game statistics (TW)}
\label{ssec:ingame}
Once we obtained the handles of the participants, we retrieved their in-game statistics via public Tracking Websites. 

\textbf{Method.} Our TW of choice is OpenDota because it provides free APIs\footnote{OpenDota API: \url{https://docs.opendota.com/}} usable to retrieve in-game statistics. We used two APIs: 
\begin{itemize}
    \item \api{player}, which, given a handle, returns some summary statistics (e.g., win/loss ratio) of the corresponding player, as well as the list of matches\footnote{For simplicity, we only considered the matches played in the previous 30 days since making each API call (i.e., from December 2019 to January 2020).} played by such a player;
    \item \api{matches}, which, given the identifier of a match (obtained from the \api{player} API), returns all information on that specific match (e.g., kills, deaths, assists).
\end{itemize}
We report in Table~\ref{tab:api} the information returned by our invoked API. Some fields are provided as lists, which include additional entries. For example, \api{matches}\_{\small chat} includes all messages exchanged by the two opposing teams during a \textsc{Dota2} match. For a detailed explanation of all fields, we refer the reader to the official documentation.

\begin{table}[!htbp]
    \centering
    \caption{Data returned by the \api{player} and \api{matches} OpenDota APIs.}
    \vspace{-0.5em}
    \label{tab:api}
    \resizebox{0.95\columnwidth}{!}{%
        \begin{tabular}{l|l||l|l||l|l}
            \toprule
            \textit{\textbf{Type}} & \textit{\textbf{Field}} & \textit{\textbf{Type}} & \textit{\textbf{Field}} & \textit{\textbf{Type}} & \textit{\textbf{Field}} \\
            \midrule
            
            num & \api{player}\_rank\_tier & num & \api{match}\_human\_players & list & \api{match}\_radiant\_team \\
            bool & \api{player}\_plus & num & \api{match}\_lobby\_type, & list & \api{match}\_dire\_team \\
            list & \api{player}\_matches & list & \api{match}\_objectives & num & \api{match}\_skill \\
            num & \api{match}\_match\_id & list & \api{match}\_picks\_bans & list & \api{match}\_players \\
            num & \api{match}\_barracks\_status\_dire & num & \api{match}\_positive\_votes & num & \api{match}\_patch \\
            num & \api{match}\_barracks\_status\_radiant & list & \api{match}\_radiant\_gold\_adv & num & \api{match}\_region \\
            list & \api{match}\_chat & num & \api{match}\_radiant\_score & list & \api{match}\_all\_word\_counts \\
            list & \api{match}\_cosmetics & bool & \api{match}\_radiant\_win & list & \api{match}\_my\_word\_counts \\
            num & \api{match}\_dire\_score & list & \api{match}\_radiant\_xp\_adv & num & \api{match}\_throw \\
            list & \api{match}\_draft\_timings & num & \api{match}\_start\_time & num & \api{match}\_comeback \\
            num & \api{match}\_duration & list & \api{match}\_teamfights & num & \api{match}\_loss \\
            num & \api{match}\_first\_blood\_time & num & \api{match}\_tower\_status\_dire & num & \api{match}\_win \\
            num & \api{match}\_game\_mode & num & \api{match}\_tower\_status\_radiant &  &  \\\bottomrule 
        \end{tabular}
    }
\end{table}

Overall, after querying the \api{players} API for each of the 484 players, we found out that our population participated in 26241 matches during the considered timeframe. Therefore, we invoked the \api{matches} API on all these entries.

\textbf{Preprocessing.}
By applying original feature engineering techniques on the data retrieved from OpenDota, we distill additional knowledge to assist in our analysis. Such techniques involve both `traditional statistics', but also our own `domain expertise' on \textsc{Dota2}.
\begin{itemize}
    \item \textit{Traditional Statistics.} The most straightforward operation involves computing some aggregated metrics on the details of each match played by a given player (e.g., average match length). We also perform some more refined operations. For instance, the \api{players} API does not directly provide the playtime trend of a given player, but such information can be computed by using the results from \api{matches}: by inspecting the dates of the matches played, we can identify, e.g., which day of the week a given player is most likely to play \textsc{Dota2}. 
    \item \textit{Domain Expertise.} By applying knowledge on the \textsc{Dota2} context, we further increase the amount of information usable for our analysis. As an example, we inspect all chat messages to determine whether players use words that are typical of \textsc{Dota2} slang (e.g., ``cd'', ``b'', ``rat'', ``smurf'', ``gank''). We provide in Appendix~\ref{app:expertise} an additional description of how we computed the features related to \api{match}\_{\small chat}.
\end{itemize}
Overall, we compute over 300 features---all of which are novel in the context of AIA.\footnote{A complete description of all our considered features is provided in our repository.} Such features identify three datasets: \smacal{P}, focused on the players, containing 484 samples, each described by 187 features; \smacal{M}, focused on the matches, containing 26241 samples, each described by 137 features; and \smacal{\overbar{M}}, containing 11117 samples and 160 features, which is a `distilled' version of \smacal{M}. In particular, \smacal{\overbar{M}} differs from \smacal{M} in two ways: First, we address the problem of the highly imbalanced distribution of \smacal{M} in terms matches-per-player (some players in \smacal{A} have only 5 matches in \smacal{M}, while others have hundreds); we thus reduce the potential bias by randomly sampling\footnote{To mitigate the effects of randomness, we create 20 versions of \sccal{\overbar{M}} and will use all of these for our experiments, averaging the results.} at most 30 matches for each player. Second, we augment the features in \smacal{M} with those derived with our domain knowledge; the intention is determining how much of an impact our intuitions (resembling those of an attacker) have on all our experiments.

\subsection{Correlation between \textsc{Dota2} in-game statistics and personal attributes}
\label{ssec:correlation}
We can now objectively determine whether a relationship exists between \textsc{Dota2} players' in-game statistics and their personal attributes. This step is crucial to provide a theoretical foundation supporting the effectiveness of AIA in this context.

\textbf{Method.} We perform a correlation analysis between our three dataset containing in-game statistics, and the dataset containing corresponding personal attributes. 
Inspired by~\cite{golbeck2011predicting}, we compute the correlation between each feature of (\smacal{P}|\smacal{M}|\smacal{\overbar{M}}), with each feature of \smacal{A}. 
To conduct a rigorous analysis, for each pair of features we compute: (i)~the \textit{statistical significance} of the correlation---measured with a $p$-value; and (ii)~the corresponding \textit{strength of the relationship}---whose measure varies depending on the chosen correlation metric. We consider two metrics~\cite{akoglu2018user}: \textit{Cramer's V} for categorical variables; \textit{Spearman's $\rho$} for numerical variables. We remark that low $p$ denotes strong significance (we set $p<0.01$ as default threshold), whereas strong relationships are denoted by high absolute values of the corresponding metric (ranging between 0 and 1).

\begin{figure}
    \centering
    \includegraphics[width = 0.75\linewidth]{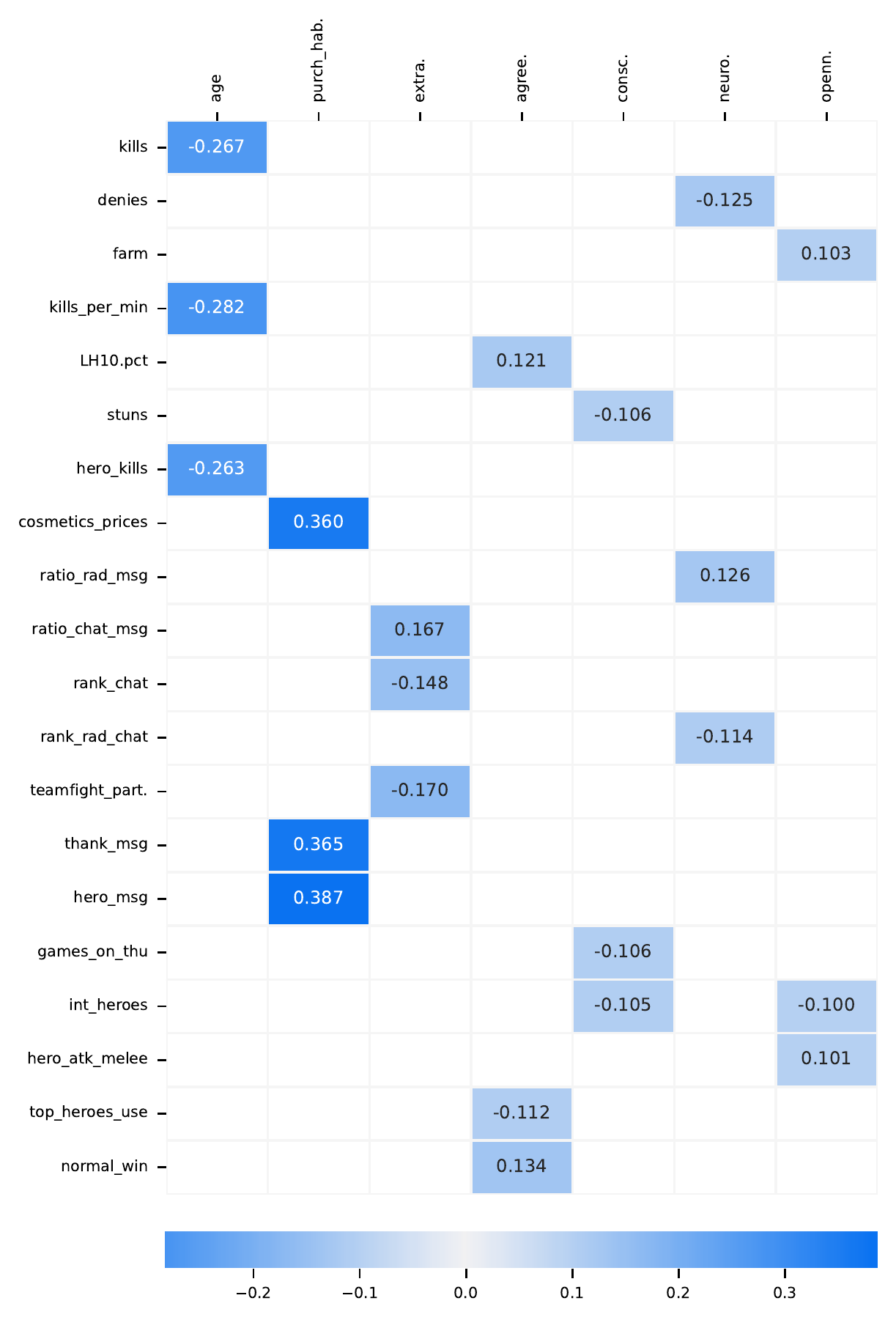}
    \caption{Top-3 Spearman's $\rho$ between \sccal{P} and \sccal{A} (at $p<0.01$). Higher absolute values denote stronger correlation, while the sign indicates the direction of the correlation.}
    \label{fig:correlation}
    \vspace{-5mm}
\end{figure}

\textbf{Results.}
We report in Fig~\ref{fig:correlation} the correlation between \smacal{P} and \smacal{A} as measured by the $\rho$ metric. For each numerical variable in \smacal{A}, we report the top-3 variables\footnote{We remark that $\rho\!>\!0.1$ is a valid signal indicator for orthogonal tasks~\cite{meyer2001psychological}.} of \smacal{P} (as measured by $\rho$), all of which obtain $p\!<\!0.01$. 
We can see that \feature{age} is correlated with \feature{kills}, probably because younger players have an aggressive playstyle. 
A strong correlation exists between \feature{purchase\_habits} and (i) \feature{cosmetics\_prices}, i.e., the money spent by a player in skins; and (ii) special messages (i.e., \feature{hero\_msg} and \feature{counter\_thank\_msg}) that can be unlocked with a paid subscription. Moreover, \feature{extroversion} is highly correlated to chat usage (i.e., \feature{rank\_chat} and \feature{ratio\_chat\_msg}); whereas \feature{agreeableness} to wins in unranked games (i.e., \feature{normal\_win}). Interestingly, \feature{neuroticism} is correlated with \feature{denies} (a unique mechanic of \textsc{Dota2}), \feature{openness} to the type of selected heroes, and \feature{conscientiousness} is low for players that play on Thursdays. 
Although not shown in Fig.~\ref{fig:correlation} (because they are categorical features), we also mention high correlation between the \feature{gender} of the player and the \feature{gender} of the most played heroes (which is common in cooperative VG~\cite{symborski2014use}); whereas the \feature{occupation} is strongly correlated to paid subscriptions.  

\begin{cooltextbox}
\textsc{\textbf{Takeaway.}}
A correlation exists between \textsc{Dota2} players' in-game data and their personal attributes. Our finding demonstrates the risk of AIA in \textsc{Dota2}. 
\end{cooltextbox}

We report in Appendix~\ref{app:correlation} an extended analysis on the correlations between (\smacal{M}|\smacal{\overbar{M}}) and \smacal{A}. Our repository includes all heatmaps.

\section{Proactive Evaluation of AIA in \textsc{Dota2}}
\label{sec:evaluation}
Our preliminary assessment provides evidence that AIA against \textsc{Dota2} can be successful. Hence, as our third contribution, we set out to proactively evaluate the impact of such AIA. To this purpose, we use the data derived from our survey (described in §\ref{ssec:survey} and §\ref{ssec:ingame}) to perform various ethical and controlled AIA.

Specifically, we find instructive to study three diverse AIA, each requiring different amounts of preparation. First, we consider the most \textit{simple} way to carry out an AIA, i.e., by using only the aggregated data of each player (§\ref{ssec:player}). Second, we evaluate the success rate of AIA that use information derived from just \textit{one match} (§\ref{ssec:one}). Third, we analyze \textit{sophisticated} AIA in which the attacker leverages all their expertise to maximize their impact (§\ref{ssec:sophisticated}).
Finally, we perform a reflective exercise by discussing the general context of AIA in light of the results achieved in research (§\ref{ssec:comparison}). We also perform
a statistical validation of all our results in Appendix~\ref{app:statistical}.

\textbf{Common Setup.}
We always adhere to our threat model (§\ref{ssec:definition}). The attacker knows the handle of one or more players, and uses such handle to retrieve in-game data from TW, which are then provided as input to an ML model for inference.
Moreover, we also assume that the attacker gathered the private attributes for training the ML model via a survey (i.e., the one described in §\ref{ssec:survey}). Indeed, as evidenced by~\cite{surveyreddit}, thousands of \textsc{Dota2} players willingly participate in game-related surveys. For ethical reasons, we do not violate our respondents' privacy by performing OSINT, or crawl their social media profiles (which are both viable means that an attacker can -- legitimately -- use to improve their AIA).

\subsection{Simple AIA (aggregated player data)}
\label{ssec:player}
The underlying principle of these AIA is that they only use the information contained in \smacal{P}, i.e., which aggregates the statistics of all matches played by any given player. Such information is simple to compute, but is lossy. For instance, the \feature{average\_match\_length} includes the duration of all matches, and inevitably leads to oversimplifications. However, due to their simplicity, such AIA are feasible to stage and it is important to assess their impact.

\textbf{Testbed.} For these experiments, we merge \smacal{P} with \smacal{A}, generating a single dataset containing 484 samples, each described by 187 features (from \smacal{P}) and associated to 9 attribute labels (from \smacal{A}).
To develop the ML model for the AIA, we consider four ML algorithms: Logistic Regression (\smamath{LR}), Decision Trees (\smamath{DT}), Random Forest (\smamath{RF}), and Neural Networks (\smamath{NN}). We validate our results through a nested stratified 10-fold cross-validation, during which we also apply feature selection and hyperparameter optimization for each considered ML model. Finally, to address the imbalance of some target attributes (e.g., \feature{age}), we apply well-known under- and over-sampling techniques~\cite{chawla2002smote, wilson1972asymptotic} (as also recommended in~\cite{arp2022and}). 

\textbf{Impact.}
We report in Table~\ref{tab:player} the results of the simple AIA. Rows denote the target attributes, whereas columns denote the considered ML algorithms; the rightmost column refers to a `Dummy' stratified classifier (simulating a random guess) which we use as baseline for comparison. Cells report the predictive macro F1-score (and standard deviation) across all our trials. 

\begin{table}[!htbp]
    \centering
    \caption{Impact of the \textit{simple} AIA (based on \sccal{P}) as measured by the F1-score. Rows report the attributes and columns our ML models (boldface denotes the best model for a given attribute).}
    \label{tab:player}
    \resizebox{0.75\columnwidth}{!}{%
        \begin{tabular}{lccccc}
            \toprule
             & \smamath{LR} & \smamath{DT} & \smamath{RF} & \smamath{NN} & \textit{Dummy}  \\\midrule
            \feature{gender} & \res{64.97}{10.9} & \res{59.71}{12.7} & \res{50.91}{5.33} & \bestres{67.24}{13.4} & \res{51.62}{10.9} \\
            \feature{age} & \res{40.47}{6.30} & \res{39.38}{8.76} & \bestres{44.08}{6.17} & \res{28.06}{7.59} & \res{32.21}{5.70} \\
            \feature{occup.} & \res{53.23}{7.22} & \res{47.44}{8.34} & \res{56.08}{7.88} & \bestres{59.89}{7.15} & \res{43.76}{9.56} \\
            \feature{purch.} & \res{32.05}{10.1} & \res{31.74}{4.53} & \bestres{34.40}{8.20} & \res{32.17}{7.19} & \res{31.20}{6.26} \\
            \feature{open.} & \res{28.94}{5.94} & \bestres{40.76}{6.80} & \res{32.6}{7.77} & \res{30.89}{7.60} & \res{29.59}{2.04} \\
            \feature{consc.} & \res{26.52}{5.65} & \res{33.87}{8.78} & \bestres{34.27}{5.60}  & \res{23.83}{8.18} & \res{33.23}{8.94} \\
            \feature{extrav.} & \res{30.15}{7.53} & \res{36.16}{5.14} & \bestres{36.49}{5.56} & \res{28.59}{5.95} & \res{32.27}{7.01} \\
            \feature{agreeab.} & \res{29.46}{6.29} & \bestres{34.11}{8.58} & \res{33.68}{6.25} & \res{24.54}{9.43} & \res{33.39}{7.35} \\
            \feature{neurot.} & \res{32.38}{6.56} & \bestres{40.76}{6.80} & \res{32.6}{7.74} & \res{31.6}{8.30}  & \res{30.07}{4.46} \\\bottomrule
        \end{tabular}
    }
    
\end{table}

From Table~\ref{tab:player}, we observe that at least one of our models always outperforms the baseline. The \smamath{NN} achieves remarkable performance (almost 70\% F1-score) to predict \feature{gender}, whereas \feature{occupation} is correctly predicted with almost 60\% F1-score. In contrast, some attributes are very difficult to predict, such as \feature{purchase\_habits} for which the performance hardly goes 3\% above the baseline.
We can conclude that such simple AIA can be effective in some cases, but real attackers can easily improve the success rate by considering additional information---as we will show in §\ref{ssec:sophisticated}.

\subsection{One-match AIA (ablation study)}
\label{ssec:one}
We now assess the effects of AIA carried out by using the statistics of just \textit{a single match}. This scenario can be considered as either a best-case or a worst-case depending on the viewpoint. Indeed, we can expect that using only one match to predict the personal attributes may yield poor results---which is a best-case for the defender. However, if such an AIA is successful, it would turn into a worst-case because the attacker can infer the private attributes with limited information (e.g., less queries to the TW API).

Moreover, we consider two attackers: an `expert' attacker that uses their \textit{domain expertise} to distill additional knowledge from the single match; and a `naive' attacker that does not do so. Hence, the results of the `naive' attacker can serve as an \textit{ablation study}, allowing to gauge the effects of domain expertise in AIA.

\textbf{Testbed.} To simulate the `naive' attacker, we merge \smacal{M} with \smacal{A}. Hence, for each of the 26241 matches in \smacal{M} (described by 137 features), we append the 9 attributes of \smacal{A}. For the `expert' attacker, we merge \smacal{\overbar{M}} (having 11117 matches, each with 160 features) with \smacal{A}, because \smacal{\overbar{M}} is augmented with \textsc{Dota2} domain knowledge. We consider the same ML algorithms as in the simple AIA (i.e., \smamath{RF}, \smamath{LR}, \smamath{NN}, \smamath{DT}). 
We then train and test ML models by adopting a split of 80:20 (such split is done on the basis of the unique players in \ftcal{M} (or \ftcal{\overbar{M}}) to avoid overfitting); we reserve 10\% of the training set for validation purposes. Finally, we repeat all our experiments 20 times to account for the random sampling of \ftcal{\overbar{M}}.

\textbf{Impact.}
We report the results in Table~\ref{tab:one}; for simplicity, we only consider the models using \smamath{RF}, because they consistently outperformed all the others. The three columns show the F1-score obtained by the `naive' (left) and `expert' (middle) attackers, as well as that of a `Dummy' classifier (right) that simulates a coin-toss. 

\vspace{-0.5em}
\begin{table}[!htbp]
    \caption{Impact of the \textit{one-match} AIA (F1-score). Columns refer to the `naive' attacker (using \sccal{M}), `expert' attacker (using \sccal{\overbar{M}}), and the Dummy (random guess). The expert attacker is always superior.}
    \label{tab:one}
    \resizebox{0.75\columnwidth}{!}{%
        \begin{tabular}{lccc}
            \toprule
            & 
             \begin{tabular}{c} {\small Naive attacker} \\ {\footnotesize (ablation study)} \end{tabular} & 
            \begin{tabular}{c} {\small Expert attacker} \\ {\footnotesize (domain knowledge)} \end{tabular} &
            \begin{tabular}{c} {\small Dummy} \\ {\footnotesize (baseline)} \end{tabular}  \\
            \midrule
            \feature{gender}  & \res{49.03}{0.18} & \res{\textbf{58.47}}{5.21} & \res{49.75}{0.55}\\
            \feature{age}  & \res{\textbf{43.72}}{2.66} & \res{\textbf{56.82}}{3.01} & \res{33.28}{0.46}\\
            \feature{occup.}  & \res{49.42}{4.56} & \res{\textbf{68.42}}{1.90} & \res{49.87}{0.89}\\
            \feature{purch.} & \res{\textbf{35.61}}{5.06} & \res{\textbf{49.71}}{3.85}  & \res{33.37}{0.53}\\
            \feature{open.}  & \res{32.26}{3.75} & \res{\textbf{43.73}}{2.96} & \res{33.48}{0.41}\\
            \feature{consc.} & \res{29.49}{3.63} & \res{\textbf{46.11}}{3.20} & \res{32.88}{0.62} \\
            \feature{extrav.}  & \res{32.33}{2.47} & \res{\textbf{46.82}}{1.96} & \res{33.25}{0.56}\\
            \feature{agreeab.}  & \res{33.62}{2.28} & \res{\textbf{45.36}}{3.37} & \res{34.09}{0.46}\\
            \feature{neurot.}  & \res{27.39}{4.78} & \res{\textbf{46.60}}{2.72} & \res{33.65}{0.58}\\\bottomrule
        \end{tabular}
    }
    \vspace{-0.5em}
\end{table}

From Table~\ref{tab:one}, we can see that the `naive' attacker cannot successfully predict 8 out of 9 attributes, because the F1-score is always comparable (or even inferior) than the Dummy classifier. The only exception is the \feature{age} attribute, for which the F1-score is 10\% superior (albeit still hardly usable). We also note that such results are inferior to those of the simple AIA (cf. Table~\ref{tab:player}). From a defender's viewpoint, these results may appear encouraging. Unfortunately, the `expert' attacker is much more successful, with 10--20\% improvements over the Dummy classifier. Notably, \feature{occupation} reaches $\sim$70\% F1-score (up from 49\%), whereas \feature{gender} almost 60\% (up from 49\%). 
Such results prove that using domain knowledge of \textsc{Dota2} substantially improves the success of AIA. What is surprising is that such AIA require the statistics of \textit{a single match} (i.e., just one API query).


\subsection{Sophisticated AIA}
\label{ssec:sophisticated}
We now assess AIA launched by a sophisticated attacker who, alongside using their domain expertise during pre-processing, exploits post-processing methods to further improve the AIA success rate. 

\textbf{Intuition.} 
We build from the one-match results of the the `expert' attacker (§\ref{ssec:one}). Then, we leverage the fact that a given \textsc{Dota2} player (i.e., the one targeted by the attacker) typically plays many matches. It is reasonable to assume that said player exhibits a \textit{stable behaviour} across all such matches. Indeed, taken individually, a single match may not capture the true behaviour of a given player, thereby leading an ML model to make a wrong prediction; however, by considering the predictions of the \textit{same} ML model to \textit{many} matches (from the same player), the stable behaviour (i.e., the desired attribute) of the targeted player is more likely to emerge. For example, a player that has `high' \feature{openness} may not show such trait in every single match; but such trait may emerge by (independently) analyzing more matches, and aggregating the results.

\textbf{Testbed.} We use the ML models trained with \smacal{\overbar{M}} using the \smamath{RF} algorithm. Then, we provide as input to such models an increasing amount of matches from the same targeted player: specifically, we consider from 1 up to 30 matches (if available), which are randomly sampled (from the test portion of \smacal{\overbar{M}}). Then, for each attribute in \smacal{A}, we take the predictions (provided as probabilities) of the ML model for all such matches, and we average all such predictions, choosing the one with the higher value.\footnote{E.g.: we want to predict the \feature{{\scriptsize occupation}} (which is binary) of a player by analyzing 4 matches. The ML model analyzes 4 matches and outputs 4 probabilities, e.g., \{0.1, 0.2, 0.8, 0.2\} (i.e., values below/above 0.5 denote employment/unemployment). We assign the class after averaging the probabilities, thereby `filtering' the noise (i.e., the 0.8).} To reduce bias, we repeat all such experiments 20 times for each different variant of \smacal{\overbar{M}}; and, we repeat the draw of the chosen matches 1000 times. 

\begin{figure}[!htbp]
    \centering
    \includegraphics[width=0.8\columnwidth]{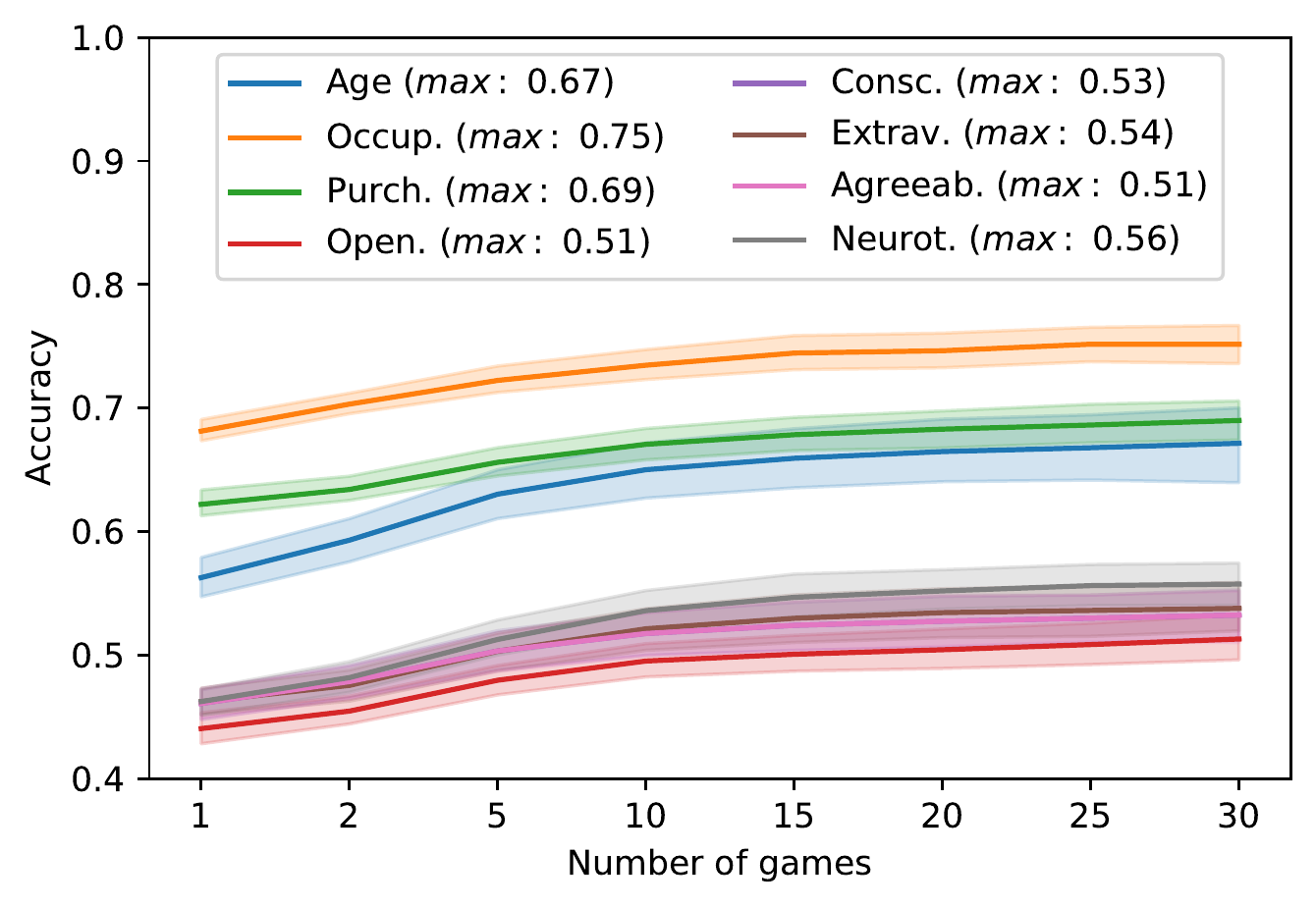}
    \caption{Impact of Sophisticated AIA. We post-process the predictions of the ML model over multiple matches of the same targeted player.}
    \label{fig:sophisticated}
\end{figure}

\textbf{Impact.}
The results of our sophisticated AIA are shown in Fig.~\ref{fig:sophisticated}, showing accuracy (y-axis) as a function of the matches analyzed by the ML model (x-axis). Lines correspond to the target attributes; shaded areas show the standard deviation. We do not report \feature{gender} because the highly unbalanced population would inflate the results. 

\vspace{-1em}

\vspace{-1em}

From Fig.~\ref{fig:sophisticated} we can see that the accuracy increases as more matches are analyzed. For example, predicting the \feature{occupation} goes from 68\% up to 75\% after 15 matches. Similarly, \feature{age} goes from 58\% up to 65\%. What makes these results concerning is that retrieving the information on such extra matches requires little effort by the attacker, because (i)~it is free and (ii)~it can be automatized.

\subsection{Reflection: AIA in research and in practice}
\label{ssec:comparison}
As a reflective exercise, we report in Table~\ref{tab:comparison} the results (according to a given `Metric') obtained by some prior works attempting to predict the same attributes considered in our paper (we exclude \feature{purchase\_habits} because it is novel). We stress that Table~\ref{tab:comparison} is not meant to be a way to compare our AIA with previous ones, since we are the first to consider the \textsc{Dota2} setting (§\ref{sssec:related}).
Moreover, past works envision (i)~different classes having (ii)~different distributions for each attribute---making any comparison unfair. 


\begin{table}[!htbp]
    \centering
    \caption{Results of prior work on AIA. Cells denote the value of a given `Metric' for each of the attributes considered in our paper.}
    \label{tab:comparison}
    \resizebox{0.95\columnwidth}{!}{%
        \begin{tabular}{c|c?cccccccc}
            \toprule
            Prior Work & Metric & \feature{gend.} & \feature{age} & \feature{occup.} & \feature{open.} & \feature{consc.} & \feature{extrav.} & \feature{agreeab.} & \feature{neurot.} \\
            \midrule
            
            Goelbeck~\cite{golbeck2011predicting} & MAE & \ours{-} & \ours{-} & \ours{-} & \ours{0.09} & \ours{0.10} & \ours{0.14} & \ours{0.11} & \ours{0.13}\\
            
            Weinsberg~\cite{weinsberg2012blurme} & AUC & \ours{0.84} & \ours{-} & \ours{-} & \ours{-} & \ours{-} & \ours{-} & \ours{-} & \ours{-}\\
            
            Al~\cite{al2012homophily} & Acc. & \ours{0.80} & \ours{0.80} & \ours{-} & \ours{-} & \ours{-} & \ours{-} & \ours{-} & \ours{-}\\
            
            Chen~\cite{chen2014effectiveness} & AUC & \ours{0.82} & \ours{0.61} & \ours{-} & \ours{-} & \ours{-} & \ours{-} & \ours{-} & \ours{-}\\
            
            Fang~\cite{fang2015relational} & Acc. & \ours{0.80} & \ours{0.73} & \ours{0.25} & \ours{-} & \ours{-} & \ours{-} & \ours{-} & \ours{-}\\
            
            Bunian~\cite{bunian2017modeling} & Acc. & \ours{-} & \ours{-} & \ours{-} & \ours{0.58} & \ours{0.60} & \ours{0.58} & \ours{0.58} & \ours{0.58}\\
            
            Yo~\cite{yo2017inference}& Acc. & \ours{0.70} & \ours{0.80} & \ours{0.70} & \ours{-} & \ours{-} & \ours{-} & \ours{-} & \ours{-}\\
            
            Mei~\cite{mei2018image} & MAE & \ours{-} & \ours{0.09} & \ours{-} & \ours{-} & \ours{-} & \ours{-} & \ours{-} & \ours{-}\\
            
            Pijani~\cite{pijani2020you} & F1 & \ours{0.83} & \ours{-} & \ours{-} & \ours{-} & \ours{-} & \ours{-} & \ours{-} & \ours{-}\\
            
            Zhang~\cite{zhang2020practical} & F1 & \ours{0.74} & \ours{0.38} & \ours{0.13} & \ours{-} & \ours{-} & \ours{-} & \ours{-} & \ours{-}\\
            
            Eidizadehakhcheloo~\cite{eidizadehakhcheloo2021divide} & AUC & \ours{0.95} & \ours{0.98} & \ours{-} & \ours{-} & \ours{-} & \ours{-} & \ours{-} & \ours{-} \\

            \bottomrule
            
        \end{tabular}
            }
\end{table}

From Table~\ref{tab:comparison}, we can see that -- from a general viewpoint -- obtaining high performance (e.g., overall \textit{accuracy}) is difficult for some attributes. However, the real threat of AIA lies in the fact that they can be customized: although precisely inferring, e.g., the \feature{age} of \textit{all} individuals among a population may be unfeasible, it is different when the objective is more specific. For instance, an attacker may want to identify just a specific group of people (e.g., children---see §\ref{ssec:feasibility}), and they can tweak their ML models for this exact purpose.

\begin{cooltextbox}
\textsc{\textbf{A positive message.}}
Our paper tackles an open issue, and our ultimate goal is to cast light on a real\footnote{The problem is real, and we demonstrated it. Our survey resembles \textsc{Dota2} population (§\ref{ssec:survey}), the statistical analysis proves the existence of correlations (§\ref{ssec:correlation}) and our evaluation shows improvements over the baselines (§\ref{sec:evaluation}).} problem---and not to aggravate such problem. Hence, for the sake of responsible research, we will now showcase only a few `practical' AIA, having near-perfect success rate.
\end{cooltextbox}
\section{Practical AIA (The true threat)}
\label{sec:practical}

Insofar, the objective of our AIA was always to infer \textit{each} class by independently considering every attribute. According to our threat model (§\ref{ssec:feasibility}), such AIA conformed to the targeted `one-to-one' category: given \textit{any} player, infer (\textit{all} of) their attributes. The results (in §\ref{sec:evaluation}), despite being arguably serious, may not induce real attackers to launch most of such AIA (aside from, perhaps, those on \feature{occupation}): some players exhibit traits that are difficult to infer.

However, attackers can also launch two other categories of AIA, which can yield `devastating' results while being surprisingly simple to carry out. As our fourth and last contribution, we now elucidate the effects of some indiscriminate `many-to-many' AIA (§\ref{ssec:indiscriminate}), and of some targeted `many-to-one' AIA (§\ref{ssec:targeted}).

\subsection{Indiscriminate `many-to-many' AIA}
\label{ssec:indiscriminate}
Let us assume an attacker whose goal is to sell the inferred attributes to the black market. Such an attacker may want to advertise their data as being ``most likely correct''; put differently, the attacker wants to ensure that the inferred information is ``unlikely to be completely incorrect'', thereby accepting some margin of error.

\textbf{Method.}
We use exactly the same setup as in the `sophisticated' AIA (§\ref{ssec:sophisticated}), where the inference is done after analyzing multiple matches. However, for these AIA, we assume an attacker who is satisfied as long as the prediction is not completely wrong. For instance, assume that a player has `high' \feature{openness} (cf. Table~\ref{tab:attributes}): we consider the AIA to be successful if the probability associated to `high' is either at the first or second place among all the possible classes (three in this case). A similar scenario describes an AIA in which the attacker wants to find, e.g., a player who is ``likely to be open'' (i.e., has `high' \feature{openness} either at the first or second place). 

\textbf{Impact.} We report the results of these AIA (after using 30 matches) in the central column of Table~\ref{tab:indiscriminate}, in which rows denote the attributes (we exclude those that only have two classes, as it would be unfair to include them); the leftmost column denotes the accuracy obtained by the sophisticated AIA (cf. Fig.~\ref{fig:sophisticated}), whereas the rightmost column denotes the improvement (as a flat difference). From Table~\ref{tab:indiscriminate}, we can see a big jump in predictive accuracy with respect to Fig.~\ref{fig:sophisticated}. For instance, inferring \feature{age} reaches 89\% accuracy, whereas \feature{purchase\_habits} goes from 65\% to 96\% accuracy. Remarkably, this method is the only one that provides usable results for \feature{agreeableness} and \feature{openness}, both with $\sim$80\% accuracy.
Despite bearing some intrinsic margin of errors (because the predicted class is not guaranteed to be the exact one), an attacker can still benefit from such imprecision, making these AIA a tangible threat.\footnote{We provide a statistical validation of these results in Appendix~\ref{app:statistical}.}

\begin{table}[!htbp]
    \caption{Indiscriminate `many-to-many' AIA (mid column). Compared to the baseline (cf. Fig.~\ref{fig:sophisticated}), the accuracy substantially increases.}
    \label{tab:indiscriminate}
    \resizebox{0.75\columnwidth}{!}{%
        \begin{tabular}{lccc}
            \toprule
            & 
             \begin{tabular}{c} {\small Sophisticated AIA} \\ {\footnotesize (30 matches)} \end{tabular} & 
            \begin{tabular}{c} {\small Indiscriminate AIA} \\ {\footnotesize (30 matches)} \end{tabular} & {\small Improvement} \\
            \midrule
            \feature{age}  & \res{67.15}{6.87} & \res{\textbf{89.15}}{4.66} & +22.00\%\\
            \feature{purch.} & \res{68.99}{3.81} & \res{\textbf{96.13}}{2.86} & +27.14\% \\
            \feature{open.}  & \res{51.30}{3.87} & \res{\textbf{77.86}}{3.39} & +26.56\%\\
            \feature{consc.} & \res{53.24}{4.88} & \res{\textbf{80.19}}{4.12} & +26.95\%\\
            \feature{extrav.}  & \res{53.78}{3.90} & \res{\textbf{81.51}}{4.40}& +27.73\%\\
            \feature{agreeab.}  & \res{50.71}{4.65} & \res{\textbf{76.84}}{5.59}& +26.13\%\\
            \feature{neurot.}  & \res{55.74}{3.88} & \res{\textbf{80.64}}{4.02}& +24.90\% \\\bottomrule
        \end{tabular}
    }
    \vspace{-1em}
\end{table}

\subsection{Targeted `many-to-one' AIA}
\label{ssec:targeted}

We now assume an attacker who wants to find players that present specific traits among a large population, e.g., finding very young players. In these cases, the attacker would train their ML models to maximize the \textit{precision} on a given class, so as to minimize the amount of false positives. Although a similar strategy inevitably leads to a reduced \textit{recall}, this is not an issue in reality: the attacker is not interested in, e.g., ``finding \textit{all} young players'' (which is an unfeasible objective), but rather ``finding a subset of those players that are \textit{guaranteed} to be young''. Such scenario is even more problematic than the previous ones, especially given that a low recall is not an issue when the population counts millions of players. 

\textbf{Targets.} We consider an attacker that is interested in identifying four ``vulnerable'' groups of players\footnote{There are over 8000 possible combinations of all our classes, and investigating all of them is clearly unfeasible and outside our scope.}. Specifically: ``very young'' (\feature{age}=\textit{13--18}), ``purchasers'' (\feature{purchase\_habits}=\textit{Rarely} $\lor$ \textit{Regularly}), and ``introverts'' (\feature{extraversion}=\textit{Low}.) Moreover, we also consider an attacker that attempts an `intersectional' AIA, wherein the targeted group conforms to two specific classes of two \textit{distinct} attributes. In this case, the attacker wants to pinpoint ``purchasers \& workers'' (\feature{occupation}=\textit{Yes}, and \feature{purchase\_habits}=\textit{Rarely}$\lor$\textit{Regularly}), which could be ideal to identify players to which advertise new products---because such players tend to make purchases, and are likely to have the economical resources for doing so (as they have a job).

\textbf{Testbed.} We adopt a similar setup of the sophisticated AIA (§\ref{ssec:sophisticated}), i.e., we use \smacal{\overbar{M}} as dataset, and evaluate the performance of our ML models as they analyze increasingly more matches of the same player, and then averaging the output probabilities. 
The crucial difference, however, lies in the problem formulation, which now reflects a \textit{binary classification} setting: the objective is predicting the targeted class, and anything outside of such class is irrelevant. To this purpose, we first merge all players that do not belong to the targeted class (i.e., the ``positive'') into a single class (i.e., the ``negative''). Then, for each target, we train a (binary) classifier by using the precision as optimization metric (whereas in the sophisticated AIA, we used the macro F1-score). 
We find the best models and hyper-parameters using a validation set having players never seen at training time, simulating that the attacker can use only data that has gathered. The good results achieved on the validation set (combined with our correlation findings described in §\ref{ssec:correlation}) suggest that the attack is feasible, and would incentivize the attackers to launch it in reality. 
Last, we evaluate the best models on the test set, having players not included in either the training or validation sets. For each targeted attribute, we repeat all these procedures five times to reduce bias and account for randomness.

\begin{figure}[!htbp]
    \centering
    \includegraphics[width=0.8\columnwidth]{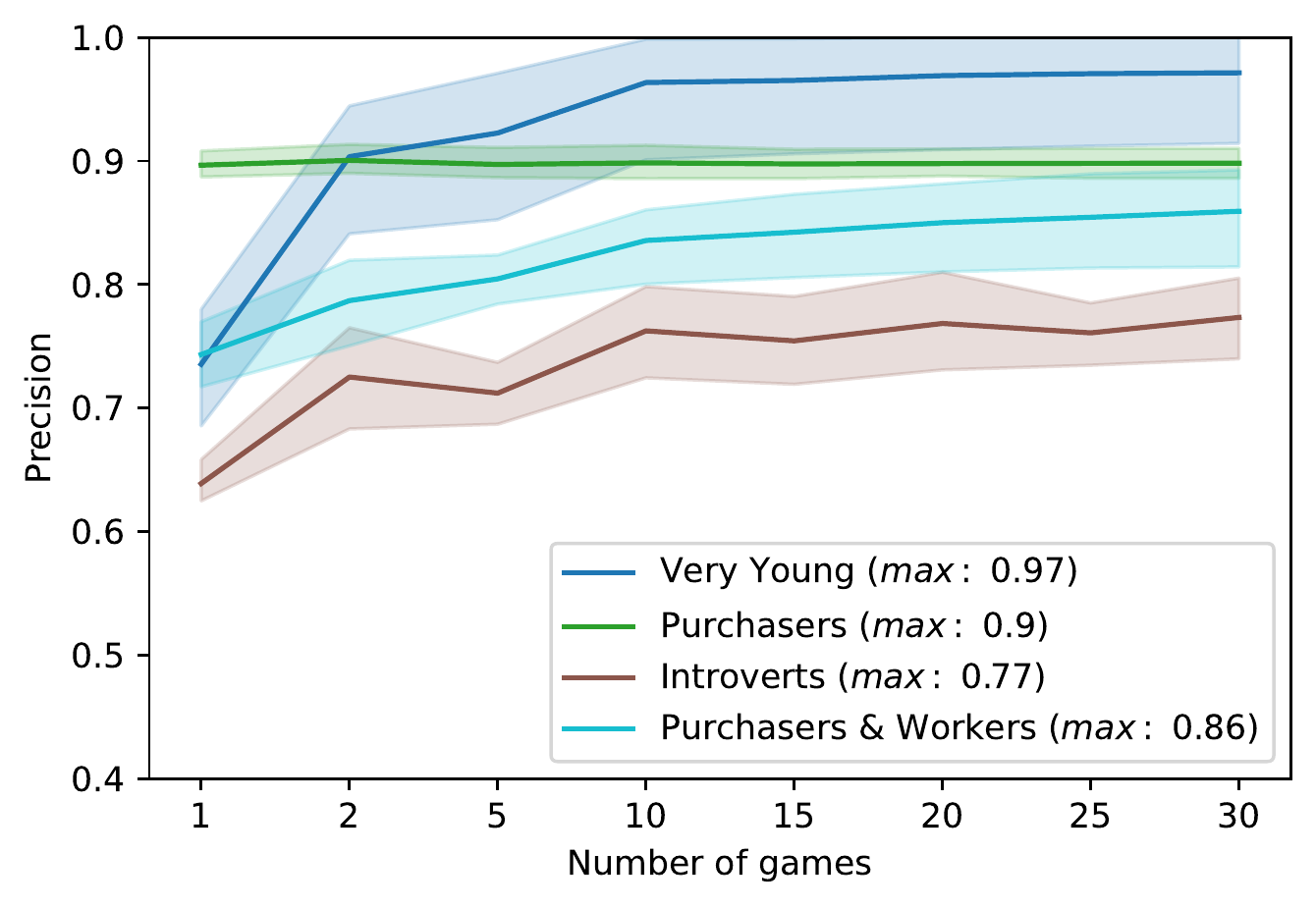}
    \caption{Targeted `many-to-one' AIA. We train our ML models by maximizing the \textit{precision} on a single targeted class. Such AIA are very effective after analyzing $\sim$10 matches for each player in the test-set.}
    \label{fig:prec}
\end{figure}

\textbf{Impact.} We report in Fig.~\ref{fig:prec} the \textit{precision} in identifying the targets as a function of the matches analyzed by the ML models. 

\vspace{-1em}
\vspace{-1em}

It immediately stands out that we obtained much `dangerous' results than in any of the previously considered scenarios. For instance, by analyzing 10 matches, our ML models can detect ``very young'' with almost perfect precision. Obviously, this comes at the cost of a low recall, which was about 47\% after 30 matches.\footnote{Roughly speaking, we detected half of the ``very young'', but with no mistakes---i.e., the ML model found $\sim$5 guaranteed ``very young'' out of $\sim$81 players in the test-set.} 
Moreover, our models ably detect ``purchasers'' after a single match, achieving a stable 90\% precision---surprisingly exhibiting also a recall of 98\% after 30 matches (not shown in Fig.~\ref{fig:prec}), suggesting that purchasing indicators are well defined, and the mistakes happened probably when users are gifted expensive items. 
The models devoted to ``introverts'' achieve 76\% precision (and 73\% recall) after with 30 matches, indicating that players belonging to this group have many characteristics in common. 
Finally, for the `intersectional' AIA focusing on ``purchasers \& workers'', the models obtain 86\% precision (and 47\% recall) after 30 matches, suggesting that roughly half of such players exhibit distinctive traits.

\begin{cooltextbox}
\textsc{\textbf{Takeaway.}}
Attackers with \textit{specific} goals can easily setup AIA that are highly successful, thereby confirming the exposure of \textsc{Dota2} players to such privacy threat.
\end{cooltextbox}

\section{Discussion}
\label{sec:discussion}

Our proactive evaluation showed that AIA can be highly successful in \textsc{Dota2}. A legitimate observation is that our experiments consider a small subset of all \textsc{Dota2} players. However, our population still allows to derive statistically significant results (see §\ref{ssec:survey}). Another observation is that we (ethically) simulated an AIA by collecting personal attributes through a survey (instead of, e.g., scraping social networks~\cite{gong2018attribute}). However, as explained in §\ref{ssec:feasibility}, \textsc{Dota2} players are willing to participate in similar surveys (even when promoted by random users~\cite{surveyreddit}). Hence, our (ethical) AIA represents a feasible scenario for an attacker, and our results are statistically significant.
Finally, there exist infinite ways in which an attacker can use the collected data to carry out AIA; yet, those considered in our paper confirm our point, i.e., that AIA are a threat to the \textsc{Dota2} playerbase.

We now discuss some possible mitigations (§\ref{ssec:countermeasures}), and explain how our threat model can be applied to other E-Sports (§\ref{ssec:extension}). 

\begin{table*}[!htbp]
\centering
\scriptsize
\caption{Overview of E-Sports VG. Numbers are taken from various sources~\cite{activeplayer,steamchart,conti2020pvp,dota_playtime,esportearn}.}
\label{tab:background}
\resizebox{1.9\columnwidth}{!}{%
\renewcommand{\arraystretch}{1.2}
\begin{tabular}{ccccccccccc}
\toprule
\multicolumn{1}{l}{} & \textit{\textbf{\begin{tabular}[c]{@{}c@{}}Release\\ Year\end{tabular}}} & \textit{\textbf{Genre}} & \textit{\textbf{\begin{tabular}[c]{@{}c@{}}Monthly\\ Players\end{tabular}}} & \textit{\textbf{\begin{tabular}[c]{@{}c@{}}Concurrent\\ Players Avg\end{tabular}}} & \textit{\textbf{\begin{tabular}[c]{@{}c@{}}Playtime\\ Avg (Hours)\end{tabular}}} & \textit{\textbf{\begin{tabular}[c]{@{}c@{}}Age Range\\ (PEGI rec.)\end{tabular}}} & \textit{\textbf{\begin{tabular}[c]{@{}c@{}}Tournament\\ Revenue\end{tabular}}} & \textit{\textbf{\begin{tabular}[c]{@{}c@{}}Exemplary \\ TW\end{tabular}}} & \textit{\textbf{\begin{tabular}[c]{@{}c@{}}Replay\\ System\end{tabular}}} & \textit{\textbf{\begin{tabular}[c]{@{}c@{}}Max Players\\ per Lobby\end{tabular}}} \\
\midrule
\textit{\textbf{\begin{tabular}[c]{@{}c@{}}League of Legends\end{tabular}}} & 2009 & MOBA & 127 M & 700 K & 832 H & 11--50 (12+) & \$93 M & lolprofile.net & Yes & 10 \\
\textit{\textbf{CS:GO}} & 2012 & FPS & 34 M & 560 K & 611H & 13--40 (18+) & \$134 M & csgostats.gg & Yes & 18 \\
\textit{\textbf{\begin{tabular}[c]{@{}c@{}}Rocket League\end{tabular}}} & 2016 & Sport & 90 M & 25 K & 315 H & 6--35 (3+) & \$18 M & rltracker.pro & Yes & 8 \\
\textit{\textbf{Fortnite}} & 2017 & Battle Royale & 270 M & 4 M & 1800 H & 6--54 (12+) & \$121 M & fortnitetracker.com & Yes & 100 \\
\textit{\textbf{PUBG}} & 2018 & Battle Royale & 510 M & 200 K & 356 H & 12--55 (16+) & \$45 M & pubg.op.gg & Yes & 100 \\
\textit{\textbf{\begin{tabular}[c]{@{}c@{}}Apex Legends\end{tabular}}} & 2019 & \begin{tabular}[c]{@{}c@{}}Battle Royale\end{tabular} & 118 M & 195 K & 91 H & 8--37 (16+) & \$10 M & apex.tracker.gg & No & 60
\\
\bottomrule
\textsc{Dota2} & 2013 & MOBA & 15 M & 450 K & 1700 H & 12--50 (12+) & \$283 M & opendota.com & Yes & 10 \\
\bottomrule
\end{tabular}
}
\end{table*}

\subsection{Countermeasures to AIA in \textsc{Dota2}}
\label{ssec:countermeasures}

Our AIA are rooted in the fact that players' in-game statistics are publicly obtainable from TW. The most obvious countermeasure would be denying public access to all such statistics \textit{from the VG itself}. Unfortunately, players are the ones (implicitly) asking for such public availability (see §\ref{ssec:ecosystem}). Alternatively, \textsc{Dota2} developers can use our analyses to make the features with stronger correlation to some attributes to be impossible to compute with public data; however, attacker are free to derive also other features---potentially with stronger correlations with (also) other attributes. 

It is hence difficult to find a `general' mitigation that preserves the functionalities of TW while ensuring players' privacy. Yet, in an attempt to reduce the feasibility of an AIA, we propose two countermeasures.
\textbf{(1)} \textit{TW could allow players to select `what content' is public.} For instance, a player can have only their last few matches to be visible by anyone. This solution has two drawbacks. First, if the statistics of \textit{other} players in the same match are visible, an attacker could still launch an AIA---albeit at a higher cost, because they need to retrieve the information from the other players (of which they need to know the handle). 
\textbf{(2)} \textit{TW could allow user to choose `who' can see their profiles.} For instance, two players could browse each other's statistics only if they are friends \textit{within the VG}---which is a different environment than the TW (e.g., Fig.~\ref{fig:screenshot} shows the friends within the TW). Such a countermeasure requires, however, a deep cooperation between TW and the VG. Alternatively, visibility can be granted \textit{upon request}. 

Unfortunately, both countermeasures impair the use of TW to learn from others players, because their matches would be hidden. The only exception are professional players, whose profiles can be public since they are less likely to be targeted AIA in the first place.

\summary{Countermeasures against AIA present tradeoffs. Our paper will hopefully inspire the search for a cost-effective solution.}

\subsection{Extension to other E-Sports}
\label{ssec:extension}

Our threat model can cover also other VG beyond \textsc{Dota2}.
Indeed, we observe that our AIA necessitates access to in-game statistics, which are mainly retrievable through TW. However, \textbf{the existence of TW is not a strict requirement}. In fact, TW elaborate statistics and replays by directly interacting with the VG itself---because it is the VG that makes such data publicly available. Therefore, an attacker could harvest these information and elaborate them autonomously. Obviously, the amount of effort required in this scenario is much higher than relying on a TW, but an AIA would still be feasible (especially for targeted `one-to-one' AIA).

Let us summarize the panorama of other E-Sports VG, for which we provide an overview in Table~\ref{tab:background}. All these VG present at least one TW akin to those of \textsc{Dota2} TW. Moreover, for all these VG, the in-game details of a player are public \textit{by default} (except for \textsc{Dota2} and CS:GO), and they often have replay system which could relax the requirement of a TW. We remark, however, that the other requirement for a successful AIA is the existence of a relationship between players' in-game statistics and personal attributes. Although there is no proof (yet) of the existence of such relationship in other contexts, we believe in its existence. In fact, many \textsc{Dota2} features can be found in the other VG. Examples are the kill/death/assist ratio, paid subscription plans and cosmetics, chat usage, or information about the play-time. Finally, we highlight that players' of some VG (e.g., Fortnite) are children, increasing the risk of AIA~\cite{fryling2015cyberbullying, groomingFort}.
\section{Conclusion}
\label{sec:conclusion}
We addresses the problem of Attribute Inference Attack (AIA) in competitive video-games (VG), with a focus on \textsc{Dota2}. We observe that \textsc{Dota2} players are naturally exposed to AIA due to the abundant in-game statistics that are publicly available. Based on this observation, we propose a threat model of AIA in \textsc{Dota2}, and (ethically) evaluate its impact. Our results demonstrate that with little preparation and domain expertise, attackers can predict the personal attributes of \textsc{Dota2} players with high success (e.g., near-perfect precision). Countermeasures to such AIA are unfeasible due to tradeoffs that would disrupt the entire \textsc{Dota2} ecosystem. 

By elucidating this subtle threat, which can affect also players of other VG, this work will hopefully inspire the development of effective mitigations (either by the VG producers, or by the TW administrators), therefore fostering an increased privacy of video gamers (who should be made aware of such risk).

\subsection*{Ethical Considerations}
\label{ssec:ethical}
Our institutions do not require any formal IRB approval to carry out the experiments described herein. Nonetheless, our survey and corresponding evaluation are all performed by adhering to the guidelines of the Menlo report~\cite{bailey2012menlo}. All interviewees were informed that their responses would be used for research purposes. Our questionnaire does not ask for sensitive data, or for private details such as name or address. We never released our dataset publicly (not even in anonymised form). All participants are also aware of the email address to contact should they be willing to have their entry removed from our dataset. Since our user-base is located in Europe, we also strictly complied with the GDPR, and all underage participants were located in countries which allowed their participation in research surveys without explicit parental consent~\cite{fra2020child}. For our AIA, we always infer the attributes that the participants of our survey willingly provided to us, hence there is no privacy violation. We do not attempt to infer personal attributes of players who did not participate in our survey (i.e., we do not collect in-game data of randomly chosen \textsc{Dota2} players, and use such data to infer their private information). The attributes we infer are non-sensitive.

\textbf{\textsc{Acknowledgement.}} We thank the Hilti Corporation for funding, and the AISec PC for the constructive feedback.



\clearpage

\appendix

\section{Extraction of Chat features}
\label{app:expertise}

We explain how we used the chat of \textsc{Dota2} for our AIA. More information is available in our repository.

\textbf{Motivation.}
Analyzing chat messages can reveal substantial information on a player.
For instance, younger players may use more slang (\feature{age}). Provocative messages can relate to nervous (\feature{neuroticism}) or energetic (\feature{extraversion}) players. Friendly (\feature{agreeableness}) players use good-behaviour messages. Efficient (\feature{conscientiousness}) players could use more tactics message, and \feature{openness} could relate to messages sent at the start of a match.
The \feature{gender} could be affected by several types of custom messages.
Finally, some hero messages must be purchased, therefore relating to \feature{occupation} and \feature{purchase\_habits}. 

\textbf{Context.}  During a match, two chat channels exist simultaneously: a \textit{team} chat, reserved for each team; and a \textit{global} chat, visible to all players. Moreover, players have the possibility to setup two  \textit{chat-wheels}\footnote{More info here: \url{https://dota2.fandom.com/wiki/Chat\_Wheel}} by choosing from a set of pre-defined messages---whose purpose is to facilitate sending of commonly used messages. In particular, each player has a \textit{general} chat-wheel (which is the same for all matches) and a \textit{hero-specific} chat wheel (which is fixed for each hero). 
Both \textsc{Dota2} and TW make public all messages sent in the \textit{global} chat, as well as all those sent with the chat-wheel (even if they are sent in the \textit{team} chat). Therefore, we use our domain expertise to extract meaningful features from both types of chat.

\textbf{Global chat.} 
We gathered lists of common English words (English is the default language for \textsc{Dota2} jargon) denoting laughs, gaming/Dota2/online slang, bad/good behavior, and provocative messages. To create such lists, we explored websites (e.g., \textsc{Dota2} forums\footnote{For instance: \url{https://dota2freaks.com/glossary/}}, urban dictionary), manually inspected thousands of match chats, and leveraged our \textsc{Dota2} expertise. Next, we counted the occurrences of such words in the player's messages. We also searched for messages containing only `?' (in \textsc{Dota2} is highly provocative), counted the number of `?', `!', and capital letters (they express astonishment or anger), the number of early-game messages (usually sent to make noise or interact with the other team), and after-kill messages (used to complain, provoke, taunt).

\textbf{Chat Wheels.} 
Messages from the chat-wheel allow to distinguish if a player is communicating in the \textit{global} (which is public) or in the \textit{team} chat (which is not publicly available). For example, a `laugh' can be sent either in the \textit{global} or in the \textit{team} chat: such difference is captured in some of our chat-wheel features. Nevertheless, such features entail tactical, laughs, deny, and good behavior messages. Moreover, we extracted which of them were `sounds', or sprays left on the ground.

\section{Additional Correlation analyses}
\label{app:correlation}
Let us expand our analysis in §\ref{ssec:correlation} with additional\footnote{A thorough description of all our correlation analyses is provided in our repository.} evidence.

Given the high number of features that describes our datasets (\smacal{M}, \smacal{\overbar{M}}, \smacal{P}), we report in Table~\ref{tab:correlations} the number of significant correlations at different $p$-values level, for both Cramer and Spearman indexes. 
From Table~\ref{tab:correlations}, we derive that many significant correlations exist for all our datasets, suggesting that ML models would be able to learn and infer private attributes starting from in-game statistics. Such a finding motivates our decision to consider AIA that use all our datasets (i.e., \smacal{P} in §\ref{ssec:player}, \smacal{M} and \smacal{\overbar{M}} in §\ref{ssec:one}).

\begin{table}[!htbp]

    \centering
    \caption{Significant Correlations at different $p$-values in our three datasets. Each column reports a personal attribute in \ftcal{A}. Rows denote how many features in each dataset (either \sccal{M}, \sccal{\overbar{M}} or \sccal{P}) achieve $p$ below the target $\alpha$ (i.e., the correlations are statistically significant).}
    \label{tab:correlations}
    \resizebox{0.99\columnwidth}{!}{%
        \begin{tabular}{ccc?ccccccccc}
        \toprule
        \textbf{\textit{Dataset}} & \textit{\textbf{Metric}} & $\alpha$ & \feature{gend.} & \feature{age} & 
        \feature{occ.} & 
        \feature{purch.} & 
        \feature{extr.} & 
        \feature{agree.} & 
        \feature{consc.} & 
        \feature{neur.} & 
        \feature{open.} \\\midrule
        \multirow{6}{*}{\textbf{\smacal{M}}} & Cram. & \textless 0.01 & 17 & 17 & 15 & 18 & 13 & 18 & 17 & 16 & 13 \\
         & Cram. & 0.05 & 18 & 19 & 15 & 18 & 14 & 19 & 18 & 19 & 14 \\
         & Cram. & 0.1 & 18 & 19 & 17 & 19 & 15 & 19 & 19 & 19 & 16 \\
         & Spear. & 0.01 & -- & 88 & -- & 51 & 44 & 52 & 22 & 70 & 36 \\
         & Spear. & 0.05 & -- & 95 & -- & 65 & 57 & 59 & 35 & 85 & 50 \\
         & Spear. & 0.1 & -- & 99 & -- & 73 & 62 & 67 & 43 & 87 & 59 \\\midrule
        \multirow{6}{*}{\textbf{\smacal{\overbar{M}}}} & Cram. & \textless 0.01 & 16 & 12 & 12 & 11 & 15 & 10 & 10 & 14 & 8 \\
         & Cram. & 0.05 & 18 & 17 & 18 & 15 & 17 & 11 & 14 & 15 & 11 \\
         & Cram. & 0.1 & 18 & 17 & 18 & 15 & 18 & 14 & 15 & 20 & 13 \\
         & Spear. & 0.01 & -- & 95 & -- & 43 & 53 & 38 & 25 & 60 & 27 \\
         & Spear. & 0.05 & -- & 104 & -- & 63 & 65 & 54 & 40 & 82 & 47 \\
         & Spear. & 0.1 & -- & 108 & -- & 69 & 73 & 64 & 53 & 90 & 58 \\\midrule
        \multirow{6}{*}{\textbf{\smacal{P}}} & Cram. & \textless 0.01 & 2 & 1 & 2 & 1 & 0 & 0 & 0 & 1 & 0 \\
         & Cram. & 0.05 & 3 & 3 & 3 & 1 & 0 & 0 & 1 & 1 & 0 \\
         & Cram. & 0.1 & 4 & 3 & 3 & 1 & 0 & 0 & 1 & 2 & 1 \\
         & Spear. & 0.01 & -- & 69 & -- & 11 & 13 & 2 & 0 & 2 & 0 \\
         & Spear. & 0.05 & -- & 97 & -- & 16 & 27 & 13 & 8 & 22 & 4 \\
         & Spear. & 0.1 & -- & 110 & -- & 26 & 47 & 26 & 16 & 44 & 14\\\bottomrule
        \end{tabular}
    }
\end{table}

\begin{figure}[!htbp]
    \centering
    \begin{subfigure}[t]{0.45\columnwidth}
        \centering
        \includegraphics[width=\columnwidth]{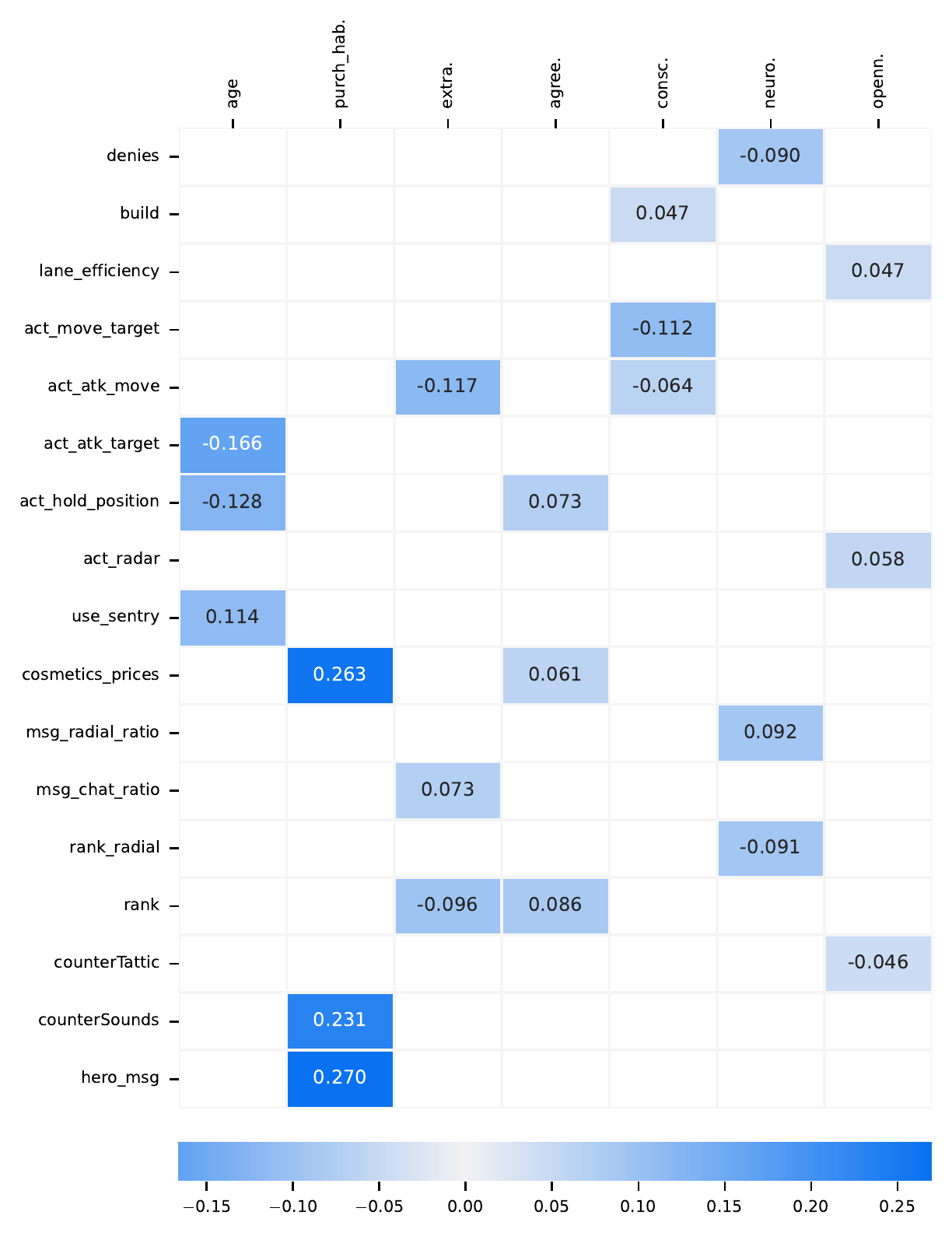}
        \caption{Correlations between \sccal{M} and \sccal{A}.}
         \label{sfig:correl_matches_M}
    \end{subfigure}%
    ~ 
    \begin{subfigure}[t]{0.45\columnwidth}
        \centering
        \includegraphics[width=\columnwidth]{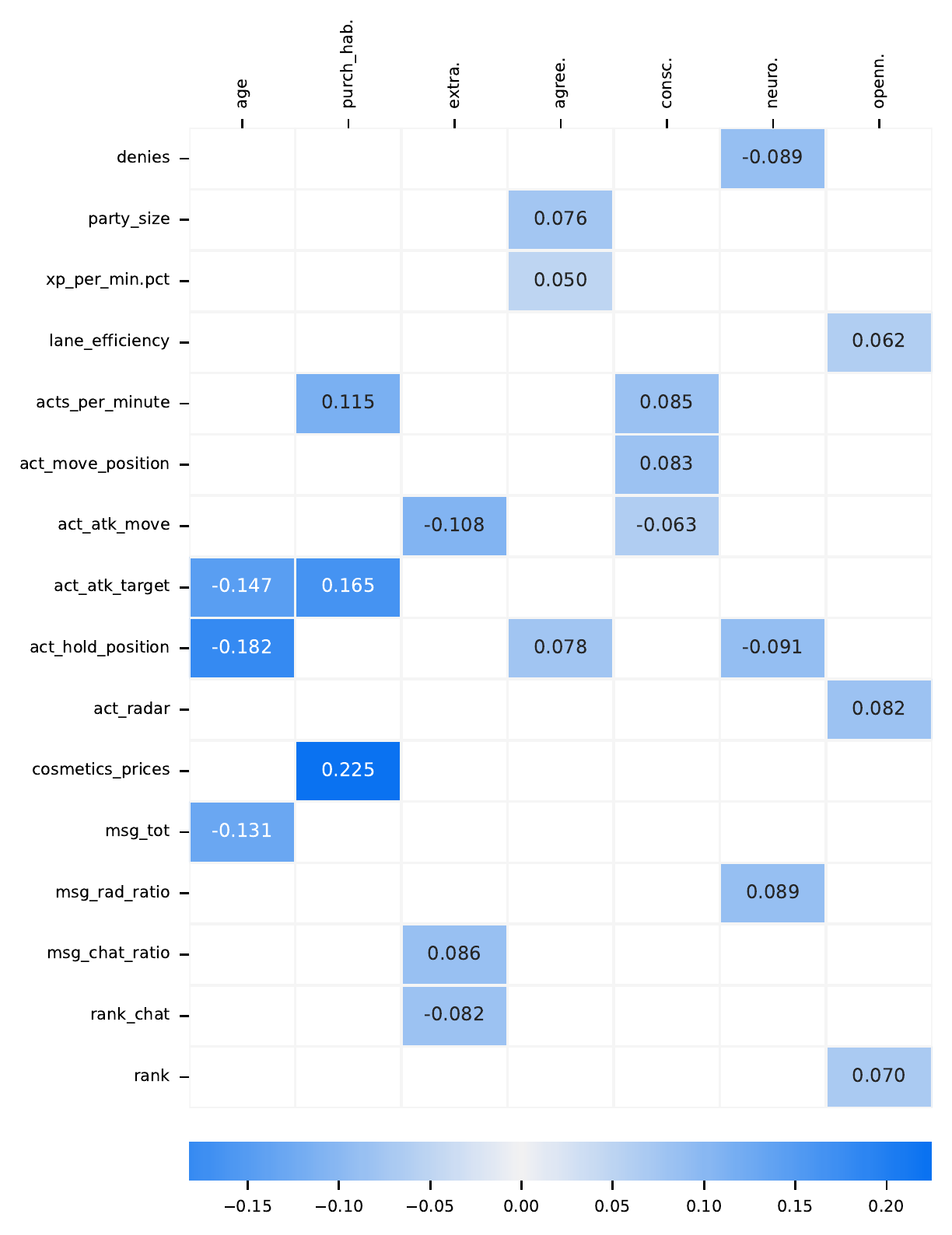}
        \caption{Correlations between \sccal{\overbar{M}} and \sccal{A}.}
         \label{sfig:correl_matches_overbarM}
    \end{subfigure}
    \caption{Top-3 Spearman significant correlation ($p$-value < 0.01)}
    \label{fig:correl_matches}
    \vspace{-1em}
\end{figure}

In Figs.~\ref{fig:correl_matches}, we report the Top-3 Spearman's correlation between \smacal{A} and both \smacal{M} (Fig.~\ref{sfig:correl_matches_M}) and \smacal{\overbar{M}} (Fig.~\ref{sfig:correl_matches_overbarM}). We observe similar strengths (e.g., compare \feature{purchase\_habits} with \feature{cosmetics\_prices}). However, personality traits tend to have low strength ($\rho\!<\!0.1$) for both of these datasets---suggesting that AIA may not be very successful at predicting such attributes.

\section{Statistical Validation}
\label{app:statistical}
We now validate the results obtained by our AIA described in §\ref{sec:evaluation} and §\ref{sec:practical}. Specifically, our goal is verifying whether our techniques achieve a performance that can be considered to be ``statistically equivalent'' to a given baseline. If such statement is found to be true, then it means that any performance difference is irrelevant; otherwise, it means that one method is better/worse than the other. 

\subsection{Methodology: two-sample Student t-test}
\label{sapp:student}
We rely on a two-sample t-test, the result of which is a $p$-value which, if superior to a given target $\alpha$, can be used to accept a given \textit{null hypothesis}.
Specifically, we set our target $\alpha$=0.05, and we set our null hypothesis as ``the technique $T_1$ is equal to the technique $T_2$''. Let us explain what $T_1$ and $T_2$ consist in by describing all of the statistical tests we perform.

\textbf{Simple AIA (§\ref{ssec:player}).} We set $T_1$ to be the performance (F1-score) achieved by the `Dummy' classifier (our baseline); whereas $T_2$ is the \textit{best} ML model for each considered attribute (i.e., the bold values in Table~\ref{tab:player}). We hence consider the corresponding values (i.e., average and std. dev.) from Table~\ref{tab:player}, and the number of samples for both $T_1$ and $T_2$ is 10 (because we use stratified 10-fold cross-validation). We perform these tests 9 times---one for each attribute in Table~\ref{tab:player}.

\textbf{One-match AIA (§\ref{ssec:one}).} Here, we perform two tests. First, we set $T_1$ to be the performance (F1-score) of the `Dummy' classifier (our baseline), and $T_2$ is the performance of the `Naive' attacker (leftmost column in Table~\ref{tab:one}). Then, we consider the same $T_1$, but consider $T_2$ to be the performance of the `Expert' attacker (middle column in Table~\ref{tab:one}). The number of samples for all these $T$ is 20 (because we repeat these experiments 20 times to account for the random sampling of \smacal{\overbar{M}}). We perform these tests 9 times---one for each attribute in Table~\ref{tab:one}.

\textbf{Indiscriminate AIA (§\ref{ssec:indiscriminate}).}
Here, we consider $T_1$ to be the performance (accuracy) of the `sophisticated AIA' (leftmost column in Table~\ref{tab:indiscriminate}), whereas $T_2$ represents the performance of the `indiscriminate AIA' (central column in Table~\ref{tab:indiscriminate}). The number of samples for both $T_1$ and $T_2$ is 20 (because we perform the draw 20 times). We perform these tests 7 times---one for each attribute in Table~\ref{tab:indiscriminate}.

\subsection{Results}
\label{sapp:results}

We report the results of all our tests in Table~\ref{tab:statistical}. Specifically, since we perform 34 comparisons in total, we report the amount of times that a given null hypothesis (in the mid-left column) must be rejected (i.e., when $p<\alpha$, mid-right column). 

\begin{table}[!htbp]

    \centering
    \caption{Statistical Validation of our results. We report the amount of tests in which the null hypothesis must be rejected (because $p<\alpha$).}
    \label{tab:statistical}
    \resizebox{0.9\columnwidth}{!}{%
        \begin{tabular}{c|c|c|c}
        \toprule
        \textbf{Table} & \textbf{Null Hypothesis} ($T_1$=$T_2$) & \# \textbf{Reject} & \textbf{Total} \\  \midrule

        \multirow{1}{*}{Table~\ref{tab:player}} & Dummy Classifier = Best Model & 5 & 9 \\
        \midrule
        \multirow{2.5}{*}{Table~\ref{tab:one}} & Dummy = Naive Attacker & 4 & 9\\ \cmidrule{2-4}
        & Dummy = Expert Attacker & 9 & 9\\
        \midrule
        \multirow{1}{*}{Table~\ref{tab:indiscriminate}} & Sophisticated = Indiscriminate & 7 & 7\\
        
        \bottomrule
        \end{tabular}
    }
\end{table}
From Table~\ref{tab:statistical}, we can see that there are cases in which our null hypothesis must be accepted, i.e., a given technique is statistically equivalent to the corresponding baseline. Unsurprisingly, these cases entail the `simple AIA' (§\ref{ssec:player}) and the ablation study (§\ref{ssec:one}).
\begin{itemize}
    \item \textbf{Simple AIA.} There are 4 cases in which ``Dummy Classifier = Best Model'', corresponding to the attributes: \feature{purchase\_habits}, \feature{conscientiousness}, \feature{extraversion}, \feature{agreeableness}. In these cases, our `simple AIA' provide a negligible performance improvement over the baseline; whereas the improvement is statistically significant for the remaining 5 attributes.
    
    \item \textbf{Ablation Study.} There are 5 cases in which ``Dummy Classifier = Naive Attacker'', corresponding to the attributes: \feature{occupation}, \feature{purchase\_habits}, \feature{openness}, \feature{extraversion}, \feature{agreeableness}. In these cases, the Naive Attacker has the same effectiveness as a coin-toss; furthermore, such an attacker is even \textit{worse} (statistically) than the Dummy classifier for \feature{conscientiousness} and for \feature{neuroticism}. The encouraging part of these results is that if an attacker could use only a single match (and is not knowledgeable about \textsc{Dota2} to derive \smacal{\overbar{M}}), then their AIA would be not very effective. 
\end{itemize}
In contrast to the above, however, our null hypothesis is \textit{always rejected} for the ``sophisticated AIA = indiscriminate AIA'', and for the ``Dummy = Expert Attacker'', thereby showing that \textbf{our `advanced' methods are always \textit{statistically superior} to the corresponding baseline}---and by a huge margin ($p<0.00001$) . 
\end{document}